\documentclass[usenatbib,useAMS]{mn2e}

\usepackage{amsmath,fleqn,graphicx,txfonts}
\usepackage{times}
\usepackage{epsf}
\usepackage[dvips]{epsfig}

\begin{document}

\label{firstpage}

\title[The formation of dSph galaxies]{A Possible Formation Scenario
  for Dwarf Spheroidal Galaxies -- \\ I: Fiducial Model}
\author[Assmann et al.]{
  P. Assmann$^{1,2}$ \thanks{E-mail: passmann@astro-udec.cl},
  M. Fellhauer$^{1}$ \thanks{mfellhauer@astro-udec.cl} ,
  M.I. Wilkinson$^{3}$ \thanks{miw6@astro.le.ac.uk},
  R. Smith$^{1}$ \thanks{rsmith@astro-udec.cl}  \\
  $^{1}$ Departamento de Astronom\'{i}a, Universidad de
  Concepci\'{o}n,  Casilla 160-C, Concepci\'{o}n, Chile \\
  $^{2}$ Departamento de Astronom\'{i}a, Universidad de Chile, Camino
  El Observatorio 1515, Las Condes, Santiago, Chile \\
  $^{3}$ Department of Physics \& Astronomy, University of Leicester,
  University Road, Leicester LE1 7RH, UK}

\pagerange{\pageref{firstpage}--\pageref{lastpage}}
\pubyear{2012}

\maketitle

\begin{abstract}
  We use numerical simulations to study a formation scenario for dwarf
  spheroidal galaxies in which their stellar populations are the
  products of the dissolution of open star clusters and stellar
  associations within cosmological dark matter haloes.  This paper
  shows that this process gives rise to objects which resemble the
  observed dwarf spheroidal satellites of the Milky Way without
  invoking external influences.  The presence of long-lived kinematic
  substructures within the stellar components of these objects affects
  their projected velocity dispersions.  We find that this in turn
  affects mass estimates based on the projected velocity dispersion
  profiles which may over-estimate the actual dark matter halo mass
  depending on the amount of substructure which is present.  Our
  models make predictions about the detailed kinematic and photometric
  properties of the dSphs which can be tested using future observations. 
\end{abstract}


\section{Introduction}
\label{sec:intro}

Dwarf spheroidal galaxies (dSph) galaxies are believed to be the
most dark matter (DM) dominated stellar systems known.  They have low
stellar content and are poor in, or entirely devoid of, gas.  They are
widely thought of as the smallest cosmological structures containing DM
in the Universe \citep{mat98,lok09,wal09} and are regarded as key
objects in the formation of larger galaxies.  According to the
$\Lambda$-CDM cosmological model, complex structures are formed
hierarchically \citep{rea06} in the potential wells of DM haloes,
having density distributions following a Navarro, Frenk \& White
(NFW) profile \citep{nav97}.  The dSph galaxies would be formed
first in small haloes, and then become involved in the mechanisms of
forming larger, complex objects.  

The dSph galaxies are characterized by absolute magnitudes
in the range $-13 \leq$ M$_{\rm V}$ $\leq-9$ \citep{mat98,Bel07}.
Their total estimated masses, considering both the stars and the DM halo,
is of the order of $10^{7}$~M$_{\odot}$ to $10^{8}$~M$_{\odot}$,
within their half-light radii.  With the advent of the Sloan Digital Sky
Survey~\citep[SDSS][]{york00}, many new faint and ultra-faint dSph galaxies
were detected around the Milky Way (MW) \citep[e.g.][and many
more]{Wil05,Bel06,Zuc06,Bel07,Wal07}.  Many of these dwarfs are less
luminous than a globular cluster (or even an open cluster) yet they
exhibit high velocity dispersions (given their luminous mass)
\citep[e.g.][]{sim07,Koc09,Geh09} and, similarly to the so-called
''classical'' dSphs,  are not rotationally supported \citep{greb03}.
Should these objects be in virial equilibrium, they are the most dark
matter (DM) dominated objects known in the universe.  They would
exhibit mass-to-light (M/L) ratios of more than a thousand
\citep[e.g.][]{sim07,Fel08,Geh09}.   

Recent works have demonstrated that dSph galaxies are DM dominated
objects \citep{kle02,kle03,kle04,mun05,mun06,wal07,sim07}. They have 
observed that the stellar velocity dispersion of the classical dSphs
is of the order of $10$~km\,s$^{-1}$ and that it remains
approximately constant with distance from the centre of the galaxy. 

There are several models that attempt to explain the origin of dSph
galaxies by considering different mechanisms.  Some of them are based
on tidal and ram-pressure stripping \citep{may07,gne99}.  In these
models, the dSph galaxies are formed due to the interaction between
a rotationally supported dwarf irregular galaxy and a MW-sized host
galaxy.  These models show that dSph galaxies tends to appear near a 
spiral galaxy, but they do not explain the presence of distant
isolated dSph galaxies, such as Tucana and Cetus.  The model proposed
by \citet{deon09} considers a mechanism known as resonant stripping
and can be used to explain the formation of isolated dSph galaxies.
Basically, these objects are formed after encounters between dwarf
disc galaxies and larger systems, in a process driven by
gravitational resonances.  There is another model that explains the 
formation of dwarf galaxies based on energy and momentum
conservation when gas-rich galaxies interact \citep{met07}.  In
this model the dSph galaxies are regarded as second-generation
objects, devoid of DM, forming in the tidal tails of gas-rich
interacting galaxies.  All the models cited above consider the
interaction between two or more galaxies to explain the formation of
dSph galaxies.   
 
There are models which consider dwarf galaxies in isolation
\citep{val08,rev09} but they usually take only a smooth gas
distribution into account and/or focus on higher masses for the
dwarfs. \citet{saw10}, for example,  perform cosmological
  re-simulations of dwarf haloes in the mass-range $2.3 \times 10^{8}$ 
  to $1.1 \times 10^{9}$~M$_{\odot}$ using an SPH code with a star
  formation recipe.  Where the density of the gas is sufficiently high
  (amongst other, more sophisticated criteria to mimic a correct
  global star formation), they convert gas particles into star
  particles, having masses of about $10^{2}$ to $10^{3}$~M$_{\odot}$.
  They do not take into account, that those masses in stars would form
  in an association or in a small dissolving cluster, because it is
  below their resolution, nor that more massive star clusters could
  form in a single star forming event.

Here, we present a different approach.  We also perform numerical
simulations of isolated galaxies (i.e.\ no other galaxy is involved)
but we distribute the newly formed stars into dissolving star clusters 
within a dark matter halo, which allow the formation of objects that
resemble classical dSph galaxies.  Our model is based on the
assumption that stars never form in isolation but in hierarchical
structures (i.e.\ star clusters) \citep[e.g.][]{lad03,lad10}.  The
star formation events range from slowly forming stars in small
clusters and associations to intense star-bursts, in gas-rich
environments, typically producing a few to a few hundred young star
clusters, within a region of just a few hundred pc
\citep[e.g.][]{whi99}.    

Those star clusters form embedded inside a molecular gas cloud and
eventually will expel their remaining gas via various feedback
processes such as stellar winds, UV-radiation and finally the onset of
supernovae \citep[e.g.][and many more and references
therein]{goo97a,goo97b,boi03a,boi03b,par08,bon11,smi11a,smi11b}. 
If the star formation efficiency (SFE) is low then the star clusters
are not able to form bound entities and instead they will disperse
their stars and dissolve.
  
In this paper, we propose that the  dynamical evolution of these star
clusters, i.e.\ their dissolution due to gas expulsion, may explain
the formation of classical dSph galaxies.  In our scenario we simulate
SCs within a DM halo , as a natural extension of the work of
\citet{fel02, wilk05, praag09}.  The star clusters form with low SFE
and, thus, are designed to dissolve inside the DM halo to form the
luminous component of the dSph galaxy.  We follow the evolution of the
star clusters within the DM halo for 10 Gyr, and then we measure the
properties of the object. 

In our models, we observe fossil remnants of the formation process
in the velocity space of the final objects of our simulations,
providing predictions for kinematic observations in the future that
could test our scenario. 

In the next section we explain our idea in more detail followed by
Sect.~\ref{sec:setup} in which we describe the setup of our
simulations.  We present the results of our fiducial model in
Sect.~\ref{sec:res} and discuss them in the final 
section.

\section{Motivation}
\label{sec:idea}

Our fiducial model is supported by two widely accepted theories.  One 
of them is the structure formation in the $\Lambda$-Cold Dark
Matter ($\Lambda$-CDM) paradigm.  In this scenario galaxies form
hierarchically in the potential well of DM haloes and small
haloes form first.  $\Lambda$-CDM cosmology with its
large N-body models of the Universe \citep[e.g. Millennium II
simulation of][]{Boy09} and the local universe around the MW
\citep[Via Lactea INCITE simulation of][]{Kuh08} predict that a galaxy
like our MW should be surrounded by many of small
DM haloes, which should/could host a dwarf galaxy as the luminous
component. 

The second theory describes the star formation inside of galaxies.  It
is now widely accepted that stars form in a clustered mode
\citep[e.g.][and follow-up publications]{lad03}.  In extreme star
bursts as in collisions of gas-rich galaxies massive star clusters or
even cluster complexes, i.e.\ clusters of massive star clusters
\citep[e.g.\ the knots in the Antennae][]{whi99} are formed.  Low
star formation rates form open clusters or associations, which have
low SFE and are more likely to dissolve.

In our scenario we simulate the DM halo of a dwarf galaxy.  In its
central region, i.e.\ where the luminous component is formed out of
the assembled baryons, we insert star clusters which have a low SFE
and therefore are designed to dissolve.  These star clusters orbit the
central region of the halo, while they are dissolving, thereby forming
the faint luminous component of the dwarf.  We have to use dissolving
star clusters to form the low-density luminous component, as the
merging of bound star clusters will only form bound compact objects
\citep[e.g.][]{fel02}.  But as the dSph galaxies are expected to only
form stars with a very low star formation rate, this assumption might
be justified \citep[see e.g.][]{bre10}. 

DSph galaxies which show globular clusters might have formed some
star clusters with a high SFE, which allowed some of them to survive
and still orbit their dwarf galaxy.

We follow the evolution of the dwarf galaxy for ten Gyr and measure
its properties at the end.

\section{Setup}
\label{sec:setup}

\begin{table*}
  \centering
  \caption{Results of fitting King, Plummer and Sersic profiles to the
  surface density data of our four realisations of the fiducial
  model (The asterisk marks the simulation we discuss separately).  The 
  last line shows the fits to the three realisations we use to take a
  mean} 
  \label{tab:sdens}
  \begin{tabular}{r|cc|cc|ccc} \hline
           & \multicolumn{2}{c}{King} & \multicolumn{2}{c}{Plummer} &
           \multicolumn{3}{c}{Sersic} \\
    number & $\Sigma_{0}$ & $R_{c}$ & $\Sigma_{0}$ & $R_{\rm pl}$ &
    $\Sigma_{\rm eff}$ & $R_{\rm eff}$ & n \\
           & [M$_{\odot}$\,pc$^{-2}$] & [pc] & [M$_{\odot}$\,pc$^{-2}$]
           & [pc] & [M$_{\odot}$\,pc$^{-2}$] & [pc] &  \\ \hline
    Sim~1  & $0.280 \pm 0.003$ & $410 \pm 10$ & $0.252 \pm 0.001$ &
    $616 \pm 9$ & $0.096 \pm 0.004$ & $500 \pm 20$ & $0.66 \pm 0.03$
    \\
    $^{*}$Sim~2 & --- & --- & $18.9 \pm 0.5$ & $61 \pm 2$ & $1.5 \pm
    0.3$ & $140 \pm 20$ & $2.1 \pm 0.2$ \\
    Sim~3  & $0.326 \pm 0.003$ & $382 \pm 8$ & $0.311 \pm 0.002$ &
    $591 \pm 9$ & $0.103 \pm 0.003$ & $540 \pm 10$ & $0.74 \pm 0.02$
    \\ 
    Sim~4  & $0.681 \pm 0.006$ & $231 \pm 5$ & $0.667 \pm 0.006$ &
    $384 \pm 9$ & $0.163 \pm 0.005$ & $450 \pm 10$ & $0.92 \pm 0.02$
    \\ \hline
%    1+2+3+4 & $14.5 \pm 0.8$ & $53 \pm 7$ & $18.4 \pm 0.5$ & $55 \pm
%    2$ & $1.7 \pm 0.3$ & $110 \pm 10$ & $2.0 \pm 0.2$ \\ \hline
    1+3+4  & $0.500 \pm 0.005$ & $251 \pm 6$ & $0.490 \pm 0.005$ &
    $420 \pm 10$ & $0.115 \pm 0.003$ & $500 \pm 10$ & $0.94 \pm 0.01$
    \\ \hline
  \end{tabular}
\end{table*}

We consider the following set-up for our simulation:
\begin{itemize}
\item The dark matter halo has a cusped Navarro, Frenk and White
  \citep{nav97} profile.  Cold dark matter only cosmological
  simulations predict all DM haloes should follow a profile which can
  be approximately parameterized in that way.  It is modeled using
  1,000,000 particles according to the recipe described in
  \citet{deh05}.  As variable parameters we use $M_{500}$, the mass
  enclosed within $500$~pc and the scale-length $R_{\rm s,h}$ of the
  halo distribution.  For our fiducial model, which is described in
  this paper $M_{500} = 10^{7}$~M$_{\odot}$ and the scale length is
  $1$~kpc.  Using a standard value for $H_{0} =
  70$~km\,s$^{-1}$Mpc$^{-1}$ and $r_{\rm vir} = r_{200}$ leads to a 
  virial radius of $12.7$~kpc for the halo ($c = 12.7$).  The value
  for $c$, i.e.\ the concentration parameter of the halo,
  falls within the predictions for dwarf haloes, which range 
  between $5$ and $20$.  The total mass of the halo out to the virial 
  radius amounts to $2.3 \times 10^{8}$~M$_{\odot}$. 
\item The luminous component is initially distributed into $N$ star
  clusters.  For our fiducial model we use $N = 30$.  This number is
  chosen arbitrarily.  We are also running simulations with smaller
  (i.e.\ more massive SCs) and larger (less massive clusters) $N$  and
  find no significant difference in our results: we are therefore
  confident that our results do not depend significantly on the choice
  of $N$ (we will present models for a range of $N$ in \citet{ass12}).
  It is claimed in \citet{bre10} that dSph galaxies had low
  star-formation rate (SFR) and low SFE.  Therefore, we form low mass
  open clusters and associations rather than massive SCs.  For
  simplicity, we neglect the SFR and start our simulations with all
  $N$ clusters forming at the same time.  However, none of our
  conclusions in the present paper depend on the fact that all the
  stars in our simulations have the same age.  Each SC is  
  modeled as a Plummer  sphere  \citep{plu11} using 100,000 particles
  \citep[ following the recipe of][]{aar74}.  The SCs have a Plummer
  radius (half-light radius) of $R_{\rm pl} = 4$~pc and a cut-off
  radius of $25$~pc.  This is similar to radii found for young SCs in
  the Antennae \citep{whi99}.  We adopt a low SFE of $30$~per cent and
  therefore the initial mass of each SC in its embedded phase is
  $5\times 10^{4}$~M$_{\odot}$.  We mimic the gas-expulsion of the SCs
  by artificially reducing the mass of each particle until the final
  mass is reached after one crossing time of the star cluster, i.e.\
  $4$~Myr.  As the mass in lost gas is negligible compared to the DM
  mass we do not take this mass further into account.  The final mass
  in stars (of all $N$ clusters) after this mass loss amounts to $4.5
  \times 10^{5}$~M$_{\odot}$, which is a typical stellar mass of one
  of the classical dSph \citep[e.g.][]{mat98}.  Note that this results
  in a particle resolution in our simulations which is slightly higher
  than reality, i.e.\ more star particles than actual stars.  
\item The star clusters themselves are distributed in virial
  equilibrium inside the halo according to a Plummer distribution.
  This distribution has a scale-length $R_{\rm pl,dist}$, which in our
  fiducial model is $250$~pc.  This takes into account that we expect
  the stars to form mainly in and around the centre of the DM halo.
  As we do not know in which virial state the SCs form with respect to
  the halo we adopt virial equilibrium for simplicity.  Also we do not
  take into account that the SCs might form in a disc-like structure
  showing angular momentum in their distribution, as we assume the
  gas-distribution on those small scales of a dSph rather supported by
  pressure than rotation.  With this assumption we differ from most of
  the previous models in which dSph galaxies are simply the
  evolutionary outcome of harassed dwarf disc galaxies, which lost
  their angular momentum because of gravitational interactions between
  galaxies.  Our model does not need any interactions between galaxies
  and therefore could explain why we see dSph galaxies in the Local
  Group far away from any major galaxies or any other dwarf. 
\end{itemize}

We simulate the cluster complex within the dark matter halo
using the particle mesh code {\sc Superbox} \citep{fel00}, which has
moving high-resolution sub-grids, staying focused on the star
clusters.  These sub-grids provide high spatial resolution at the places
of interest.  In {\sc Superbox} each simulated object has two
levels of high resolution sub-grids. The highest resolution grid
has a resolution (i.e. cell-length) of $67$~pc for the dark matter
halo and $0.8$~pc for the star cluster and covers the central area of
the halo or the SCs completely respectively.  The medium resolution
grid has a cell-length of $333$~pc for the DM halo and $166.6$~pc for
the SCs.  Finally the outermost grid covers the complete area after
the virial radius of the dark matter halo with a resolution of
$1.6$~kpc.   The time-step is fixed at $0.25$~Myr to resolve the
internal dynamics of the SCs.

As the SCs dissolve immediately, star particles are collision-less and
two-body relaxation effects are unimportant, i.e.\ we are able to use
a fast particle-mesh code.  A particle-mesh code naturally neglects
close encounters between the particles (which here are rather
representations of the phase-space than actual single stars) and is
therefore called collision-less.  That the particles are phase-space
representations makes it (in our case) possible to actually use more
particles than actual stars.  With the same reasoning we can model the
DM halo without using an actual number of possible DM particles. 

\section{Results}
\label{sec:res}

\begin{figure*}
  \begin{center}
    \epsfxsize=8.5cm
    \epsfysize=8.5cm
    \epsffile{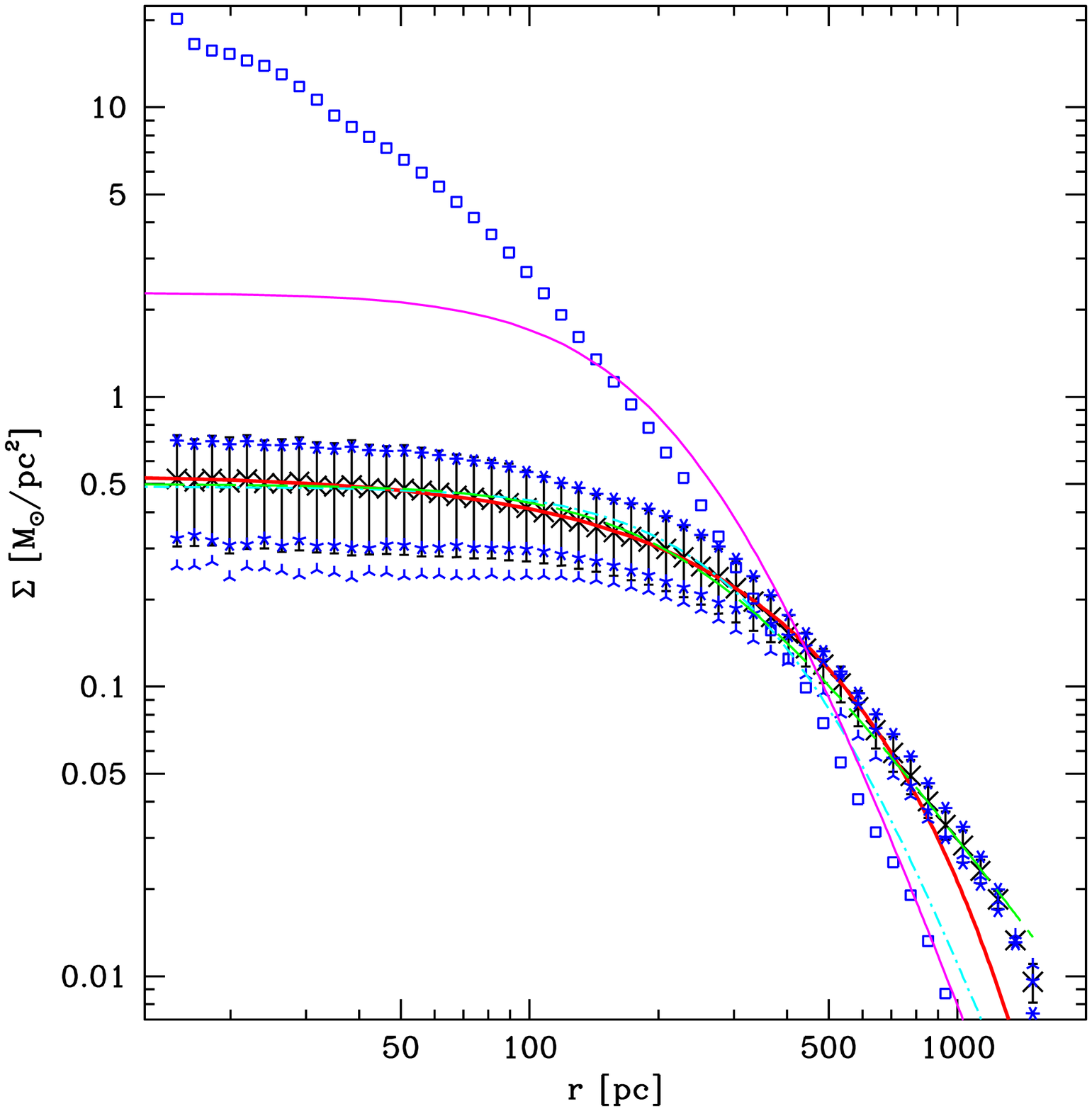}
    \epsfxsize=8.5cm
    \epsfysize=8.5cm
    \epsffile{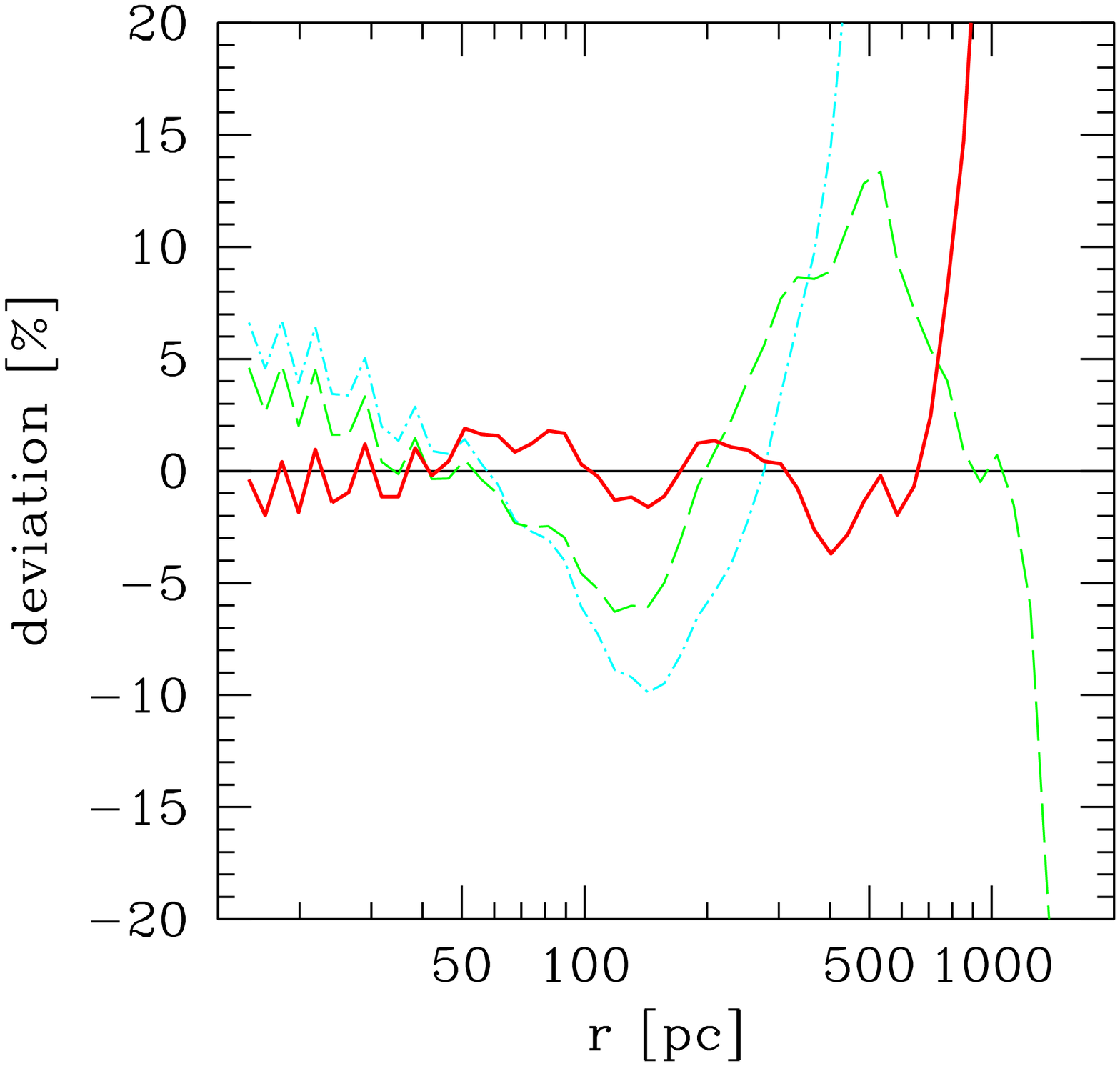}
    \caption{Left: Surface brightness profiles for the fiducial models
      of the dSph galaxy: The (blue) stars are realisations 1
      (tri-pointed), 3 (five-pointed) and 4 (six-pointed) which we use
      to form the mean values.  The open squares represent the results
      of realisation 2 which formed a central cusp, which we are not
      using.  The data points are in themselves mean values, formed
      out of three different sight-lines along the Cartesian
      coordinate axes, as we do not know the actual orientation under
      which we would see the objects.  The (black) crosses are the
      mean values with error-bars obtained out of the data-points
      shown from the three realisations used.  The (red) solid line
      shows the Sersic profile fitted to the mean data, the long
      dashed (green) line represents a King fit without tidal radius
      and finally the dashed-dotted (blue) line is a Plummer fit. 
      Finally the magenta curve shows the initial distribution of the
      star clusters.  Right: Shows the deviation of the data points from
      the fitted profiles.  It shows clearly that the Sersic fit
      represents the data best until a radius of about $700$~pc.} 
   \label{fig:sdens}
 \end{center}
\end{figure*}

We use $30$ randomly placed (according to the Plummer distribution
explained above) SCs in our simulations.  Some of the results may
therefore suffer from this low-number under-sampling of the
distribution.  In other words, two realisations of the same initial
conditions can lead to very different results, if the SCs are placed
at different positions and different orbits, according to the same
underlying distribution. 

Therefore we run four different random realizations of our fiducial
model and measure the radial profiles in all of the simulations.  We
then calculate a 'mean' profile to present in this paper.
Furthermore, each simulation model may look different, when viewed
from another direction, even though we simulate our models in
isolation and therefore no preferred direction exists.  To accommodate
the fact of the unknown sight-line we calculate for each simulation
the profiles as seen from three different sight-lines (along the
Cartesian coordinate axes) and calculate a mean value as well.  I.e.\
the profiles shown in this paper are not only an average over three
sight-lines but also of different random realisations of the same set
of parameters.  With this technique we account for the effects of
low-number random realisations. 

We exclude one model which had a chance merger of still undisrupted
SCs right at the beginning of the simulation in the centre and
therefore formed a cuspy luminous profile, as the stars could not
leave the deep potential well of the centre of the DM halo.  All other
SCs in this and all other realisations are dissolved as intended.
This particular simulation shows a dSph galaxy with a nucleus.  As we
do not see any  MW dSph with a nucleus we exclude this model from the
calculation of the mean values but include it in the paper to show
what could happen in a rare case.  We emphasise that this model is not
a numerical error but rather constitutes a rare event (given our
initial conditions), which does not appear to have been realised 
with the MW dSph population we are attempting to model (with the
possible exception of Sagittarius).  Even though our setup for the
fiducial model is tailored to have all SCs dissolving, there is still
a chance that SCs survive the disruption.  We will return to this
issue in the follow up paper, in which we investigate a wider
parameter space.

For the two-dimensional plots we focus on just one of our random
realisations (No.~1). 

A potentially important prediction of our model is that we need
slightly less DM to explain the properties of the classical dSphs
than is usually invoked.  This result is already implicit from the
setup section as our fiducial model is a model with only
$10^{7}$~M$_{\odot}$ of DM within $500$~pc, in contrast to the
observational values for the classical dSph which are of the order 
of $10^{7}$~M$_{\odot}$ ($0.5$--$1.5$) as well, but within a radius
of only $300$~pc \citep[e.g.][their figure~4]{wal09}.  We show in
the following sections that our fiducial model does in fact match
the observational data we find in classical dSph galaxies.

\subsection{Surface brightness distribution}
\label{sec:sdens}

\begin{figure*}
  \epsfxsize=5.8cm
  \epsfysize=5.8cm
  \epsffile{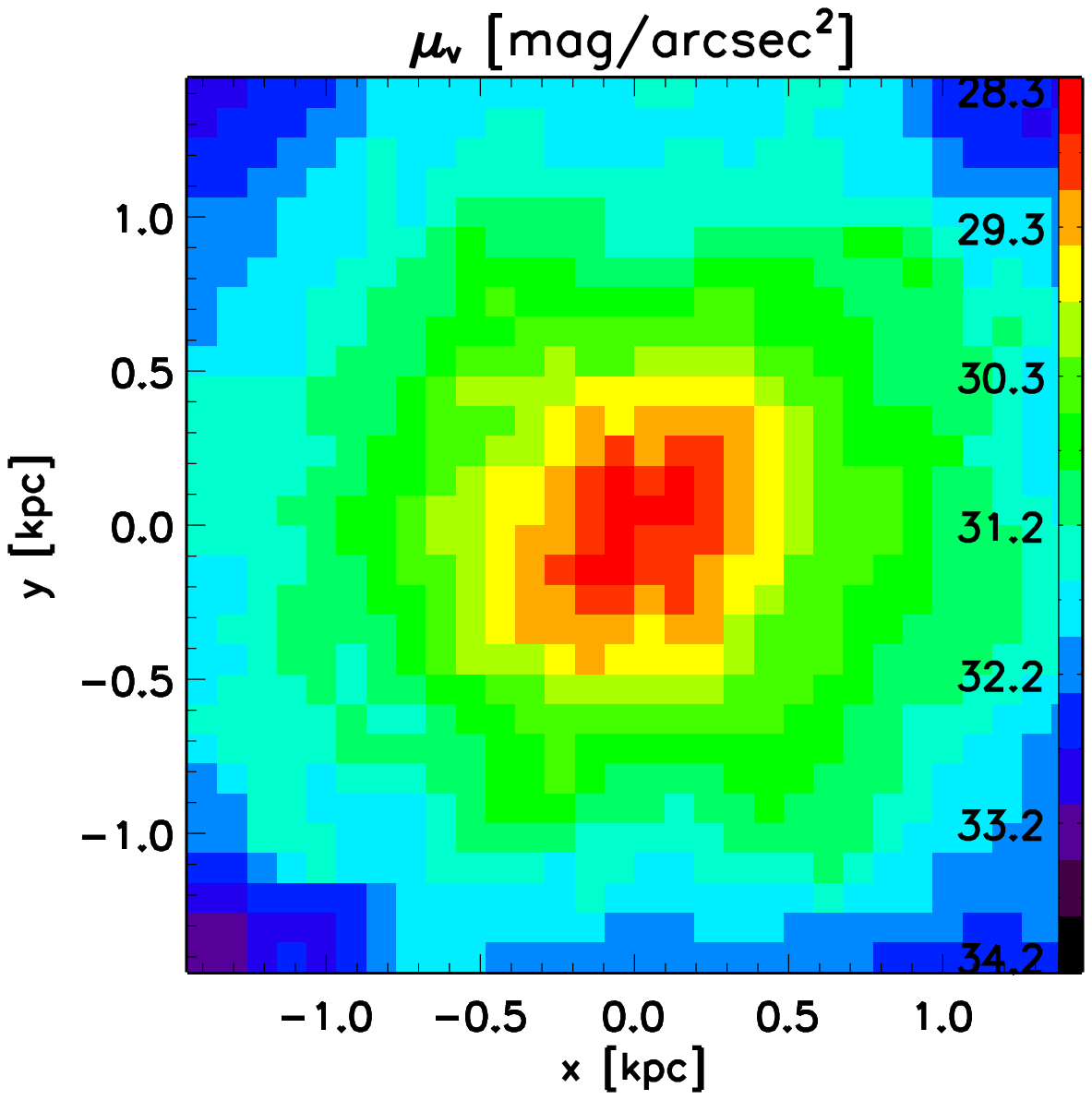}
  \epsfxsize=5.8cm
  \epsfysize=5.8cm
  \epsffile{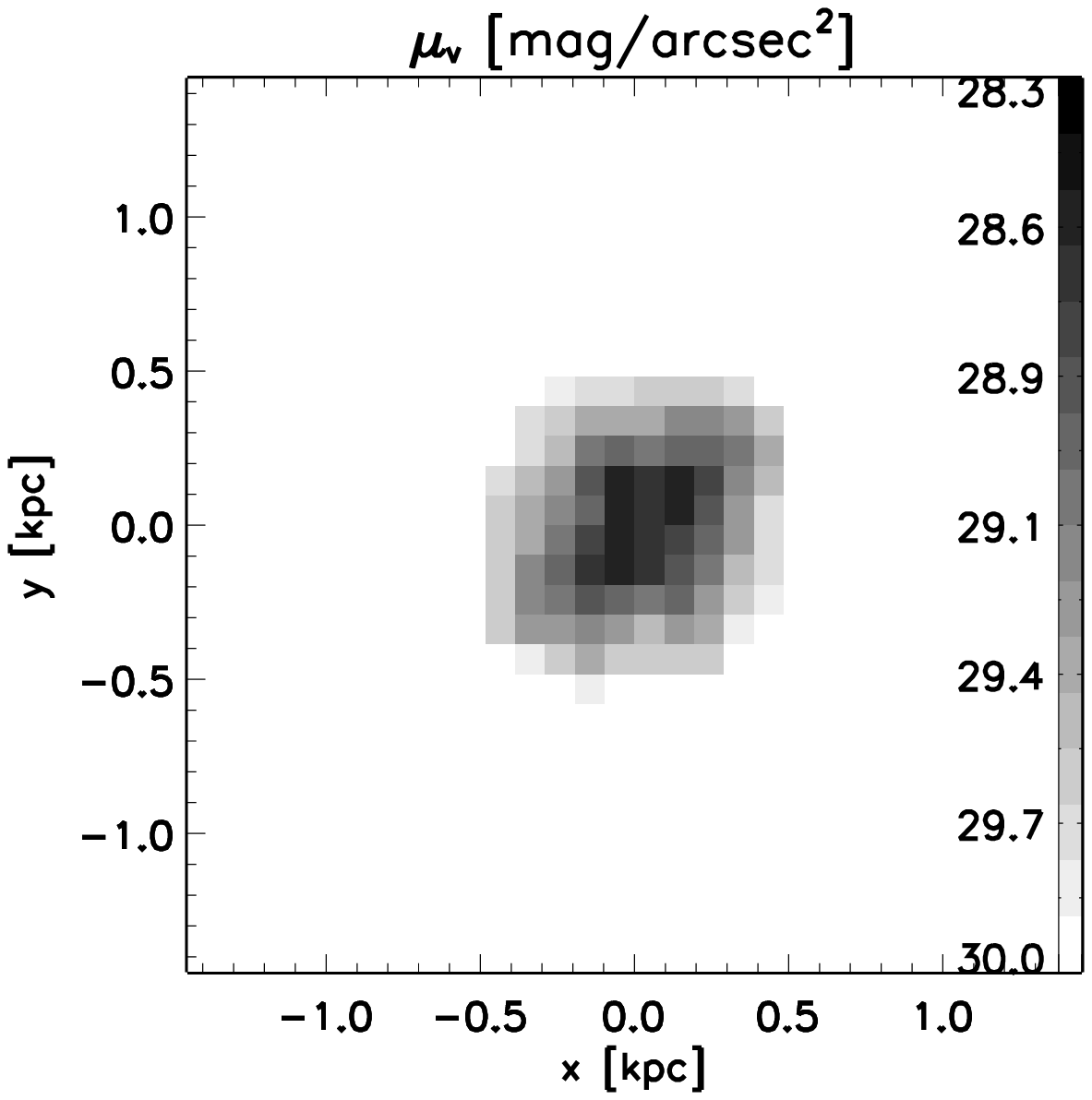}
  \epsfxsize=5.8cm
  \epsfysize=5.8cm
  \epsffile{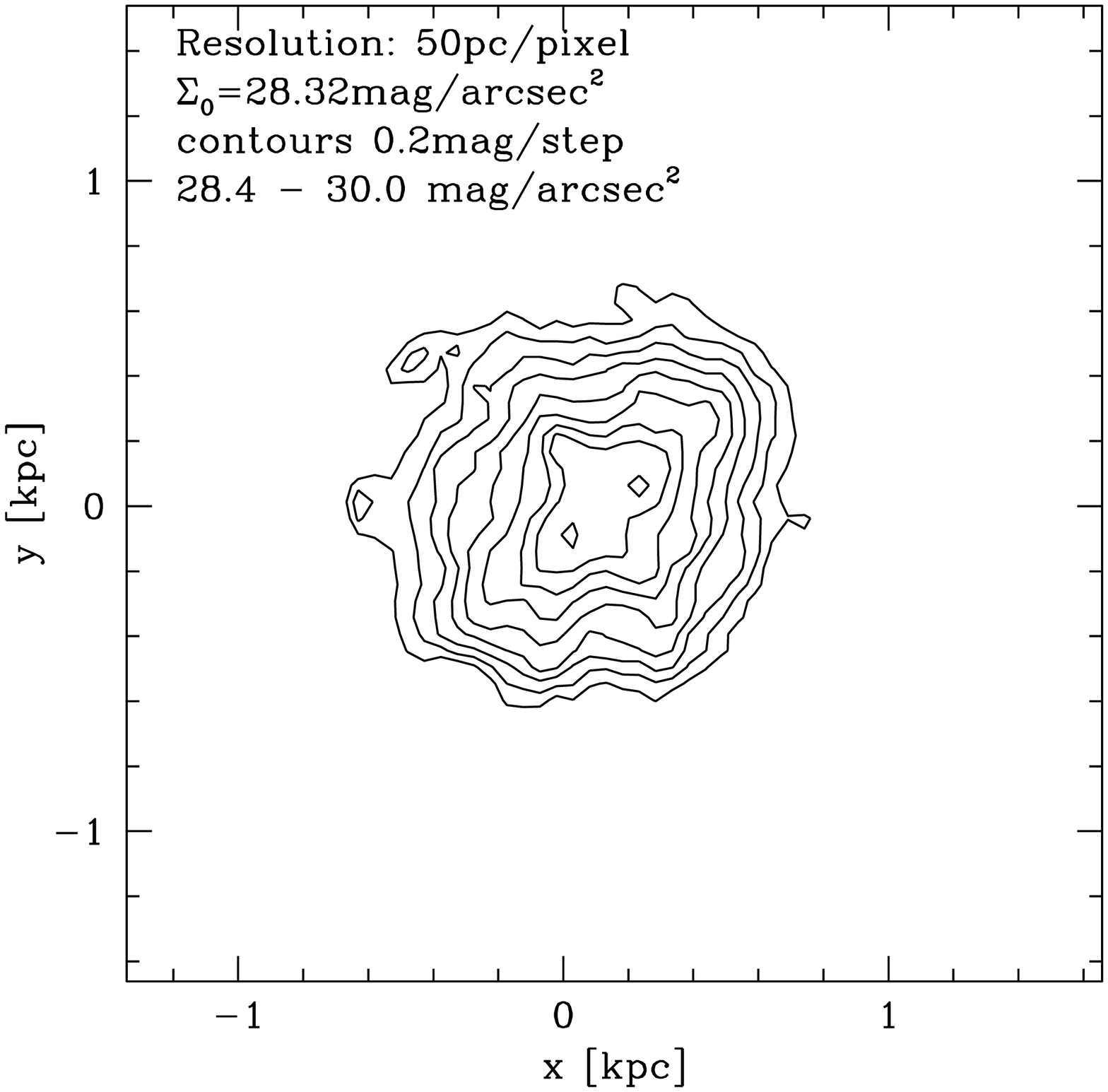}
  \epsfxsize=5.8cm
  \epsfysize=5.8cm
  \epsffile{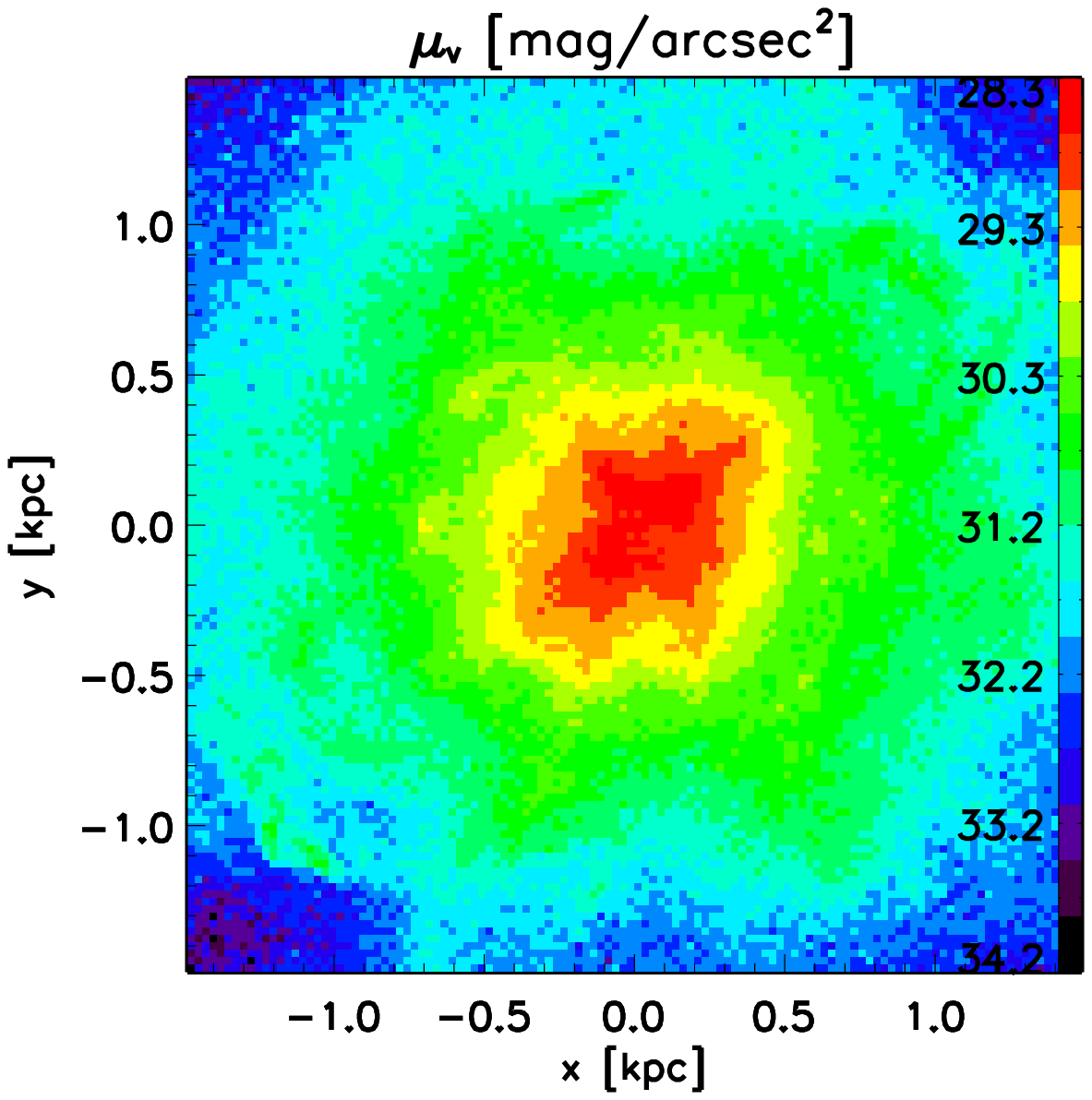}
  \epsfxsize=5.8cm
  \epsfysize=5.8cm
  \epsffile{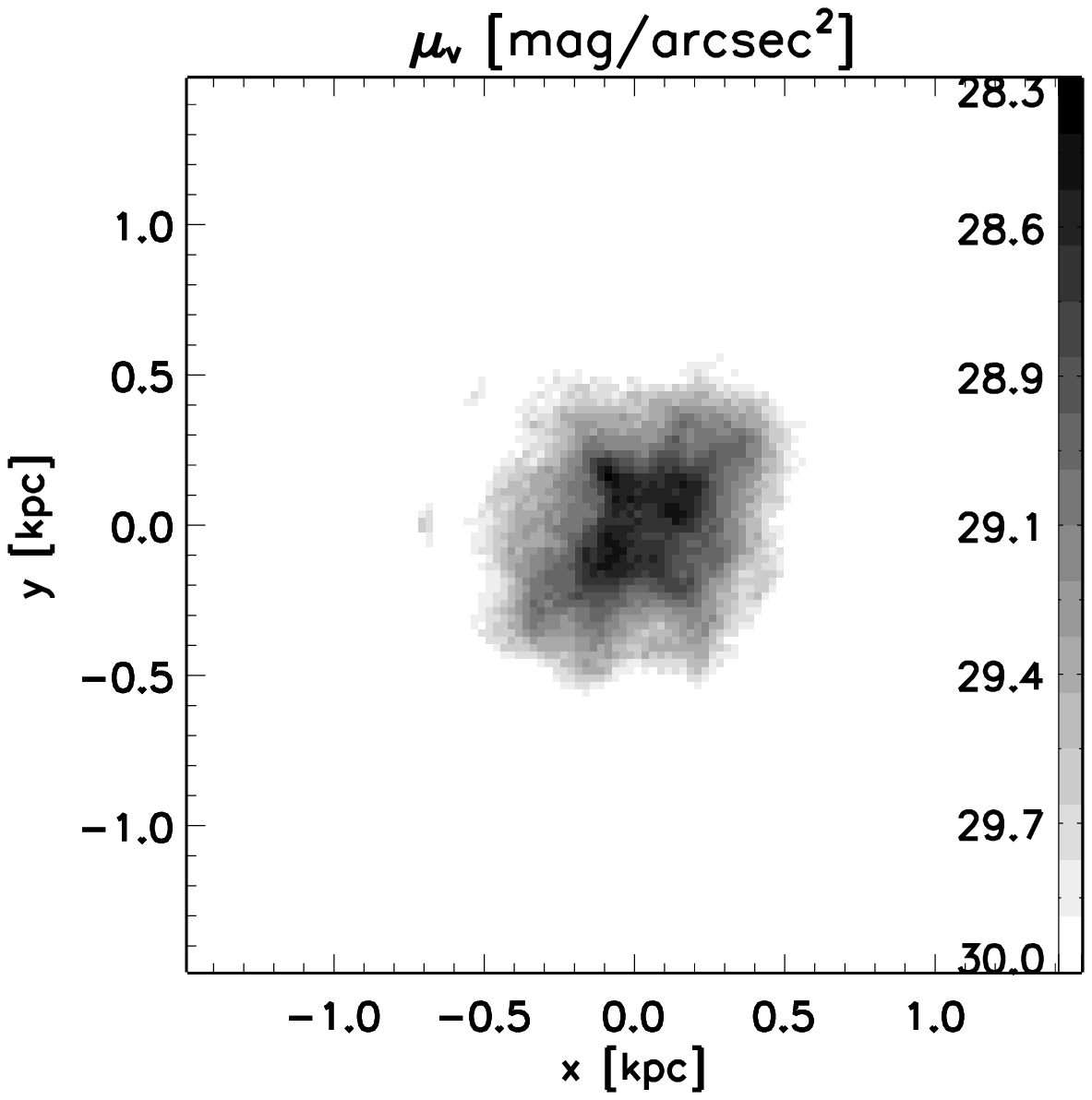}
  \epsfxsize=5.8cm
  \epsfysize=5.8cm
  \epsffile{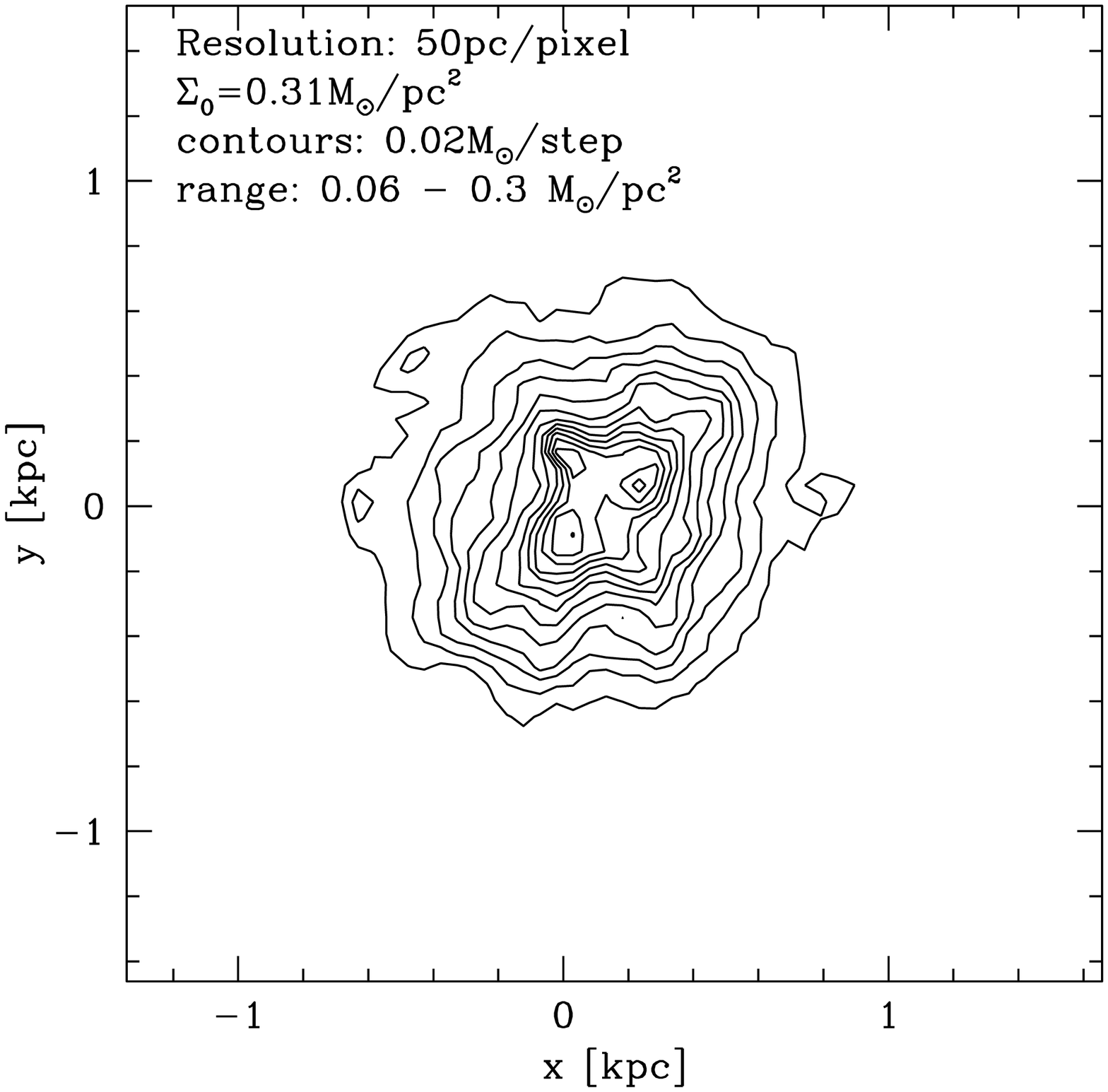}
  \centering
  \caption{Two dimensional surface brightness pixel maps and contour
    plots of our fiducial model.  The upper left and middle panels
    show the object with a resolution of $100$~pc per pixel.  The
    lower left and middle panels show the same model with a high
    resolution of $25$~pc per pixel.  With these two resolutions we
    encompass the range of resolutions of the contours produced by
    \citet{irw95}, which ranges from $24$~pc per pixel for Draco to
    $89$~pc per pixel for Fornax.  In the left panels we do not apply
    a magnitude cut and show even the faintest structures visible with
    our high particle resolution, i.e.\ in the low resolution case
    features ranging from $28.53$ to $33.23$~mag\,arcsec$^{-2}$ and in
    the high resolution case the range is from $28.32$ to
    $34.12$~mag\,arcsec$^{-2}$, while the middle panels show the model
    with a magnitude cut of $30$~mag\,arcsec$^{-2}$ applied.  The
    colour-bars are normalised to values between $28.3$ and
    $34.2$ (30.0 respectively)~mag\,arcsec$^{-2}$.  In the right
    panels we show contour plots of the fiducial model, now with an
    intermediate resolution of 50~pc per pixel. The top panel shows
    magnitude contours (logarithmic scale) while the bottom panel
    shows mass per pixel contours (linear scale).  The lower contours
    are better for the comparison with \citet{irw95}, as they plot
    star-counts as sigma-levels above the background (linear scale).} 
\label{fig:smooth}
\end{figure*}

In Fig.~\ref{fig:sdens} we show the surface brightness profiles of all
realisations of our fiducial model.  As we do not know the actual
line-of-sight we take for each realisation a mean value out of the
three lines-of-sight along the Cartesian coordinate axes.  Furthermore
and as stated before, to avoid small number statistics we plot as the
large crosses with one-sigma error-bars the mean values taken from
three of our simulations (i.e.\ data-points are means from $3\times3$
values).  We fit these mean values with a King, Plummer and a Sersic
profile.  As we do not expect a tidal radius inside a DM halo we use
eq.~13 from \citet{kin62} and fit a central surface density of 
$\Sigma_{0} = 0.500 \pm 0.005$~M$_{\odot}$\,pc$^{-2}$ which translates
(using a mass to light ratio for the stellar component of $1.0$) into
$27.2$~mag\,arcsec$^{-2}$ and a core radius of  $R_{c} = 251 \pm
6$~pc.  The Plummer fit leads to a central surface density of
$\Sigma_{0} = 0.490 \pm 0.005$~M$_{\odot}$\,pc$^{-2}$ and a Plummer
radius (which corresponds to the half-light radius) of $R_{\rm pl} =
420 \pm 10$~pc.  Finally, the Sersic fit 
\begin{eqnarray}
  \label{eq:sersic}
  \Sigma(R) & = & \Sigma_{\rm eff} \exp \left( -b_{n} \left[ \left(
          \frac{R}{R_{\rm eff}} \right)^{1/n} - 1 \right] \right) \\
  b_{n} & = & 1.9992 n - 0.3271 \nonumber
\end{eqnarray}
gives a surface density at the effective radius of $\Sigma_{\rm eff} =
0.115 \pm 0.003$~M$_{\odot}$\,pc$^{-2}$ or $28.8$~mag\,arcsec$^{-2}$,
an effective radius of $R_{\rm eff} = 500 \pm 10$~pc and an index $n =
0.94 \pm 0.01$, i.e.\ almost an exponential profile.  This matches the
profiles of the classical dSph of the MW and Andromeda quite well
\citep[e.g.][]{irw95,mat98}, even though our models seem to be a bit
on the lower luminosity side.  However, this is a consequence of our
chosen initial conditions, we would clearly expect higher luminosities
if we insert more stars in the form of more clusters.  The exact
parameter values for all realisations can be found in
Tab.~\ref{tab:sdens}.  

But it is clearly visible that the final configuration of our luminous
component has nothing in common with the initial set-up of our
models.  A Plummer profile fits not only very poorly to our model data
but also has a Plummer radius of about almost twice the initial
value.  Furthermore, we see that our objects are well fitted by
exponential profiles and therefore we conclude that the profile of the
initial distribution is of minor influence on the final profile.

In Fig.~\ref{fig:smooth}, we show two-dimensional surface
brightness pixel maps of the first of our simulations of our fiducial
model using colour pixels (pixel values are indicated with the colour
bar).  In the top left panel (shown at resolution of $100$~pc), we can
observe a smooth object with values for the surface brightness between
$28.2$ and $33.5$~mag\,arcsec$^{-2}$.  In the region of very low
surface brightness we see a hint of arms and spikes, which are
remnants of the formation process.  They are visible only because we
have ideal conditions in our simulations.  We do not have any
foreground, background or halo star contamination and we 'see' all the
stars of the dwarf galaxy, while observers sometimes only see the
upper part of the main sequence and evolved stars (e.g.\ red giants).
In fact our mass resolution is even better than reality as the
particles in our simulations can have arbitrary mass (here less than
one star) and therefore we have an over-sampling in our simulations.
If we apply a magnitude cut of $30$~mag\,arcsec$^{-2}$, roughly the
intensity limit which a telescope can detect \citep{maj05} (top middle
panel), we see that these faint structures disappear.  

The bottom left and bottom middle panels show our dSph galaxy with a
much higher resolution of $25$~pc.  Here we really take advantage out
of our over-sampling.  Again the left panel shows even fainter
structures, while the middle has a magnitude cut of
$30$~mag\,arcsec$^{-2}$.   

The central part still looks rather smooth but we see hints for
deviations from smooth ellipses.  In the middle panel one could suspect
that the actual centre of density is not in the centre of the galaxy
but rather splits up into two or three components, which are slightly
off-centre.  Those inner structures could resemble the sub-structures
of Ursa Minor as reported by \citet{kle03}.  

In the inner parts violent relaxation processes triggered by the cuspy
DM halo have erased and smoothed out most of the features.  We see in
our simulations that star clusters which are on rather radial orbits 
spread their stars smoothly in the central region, leaving no trace of
their initial orbit.  Star clusters on rather circular orbits have the
chance to spread their stars in ring-like structures, which may
survive even after 10~Gyr.  The further out these rings are, the
longer is the local relaxation time and higher is the chance to
observe them even after 10~Gyr.  But even in the central part
the dSph galaxies have very long relaxation times so some unrelaxed
parts might be expected. 

In the outer parts, the very-low density regions
($>30$~mag\,arcsec$^{-2}$), where relaxation times are longer we
clearly see arcs and spikes.  These stem from our formation process in
which some star clusters orbit out to large radii thereby slowly
dissolving and losing their stars.  Those arcs are clear sign of
stars still moving on the former orbits of the star clusters we used
to build the galaxy.  In the outer parts these structures survived the
$10$~Gyr of evolution unharmed.  We see trails similar to the ones
around major galaxies which have undergone mergers.  But in a dSph
their surface brightness is too low to be detected with today's
observations. 

The choice of our resolutions represents on the high-resolution side 
(lower row) the resolution we have in our grid-code (actual
grid-resolution is $16$~pc).  Furthermore with the choice of $25$ and
$100$~pc we encompass the resolutions shown in \citet{irw95}, which
range from $24$~pc per pixel for Draco to $89$~pc per pixel for Fornax.

But how can we compare the results of our simulations with the actual
observations of \citet{irw95}?  In the Fornax galaxy, we would expect
to detect deviations of smooth contours more easily, as Fornax has more
stars and therefore sub-structures could be stronger than noise.
Unluckily Fornax is far away and we do not have sufficient resolution.
This might be the reason why Fornax rather looks smooth like the
top-right panel of our figure.  On the other hand the observations of
Draco have sufficient resolution but Draco is less luminous (a factor
of ten in star counts at all radii), so there are not enough bright
stars to distinguish features like in our high-resolution panels from
noise.  Still, in fig.~1 of \citet{irw95} there is a hint of a
deformation in the contours of Draco.

For a better comparison we include the two right panels which show
contours of our model with a medium resolution of 50~pc per pixel.
The top panel uses again a magnitude scale while the bottom panel has
a linear mass-scale.  As \citet{irw95} also use a linear scale
(sigma-deviations of counted stars above the background) the lower
right panel gives the best match to the figures in that observational
paper.  We see bumps and dents in the contours like we see in Carina
(dents) or in Sextans (bumps).  We also see that the central part has
two density peaks, similar to what we see in Ursa Minor.

\begin{figure*}
  \begin{center}
    \epsfxsize=8.5cm
    \epsfysize=8.5cm
    \epsffile{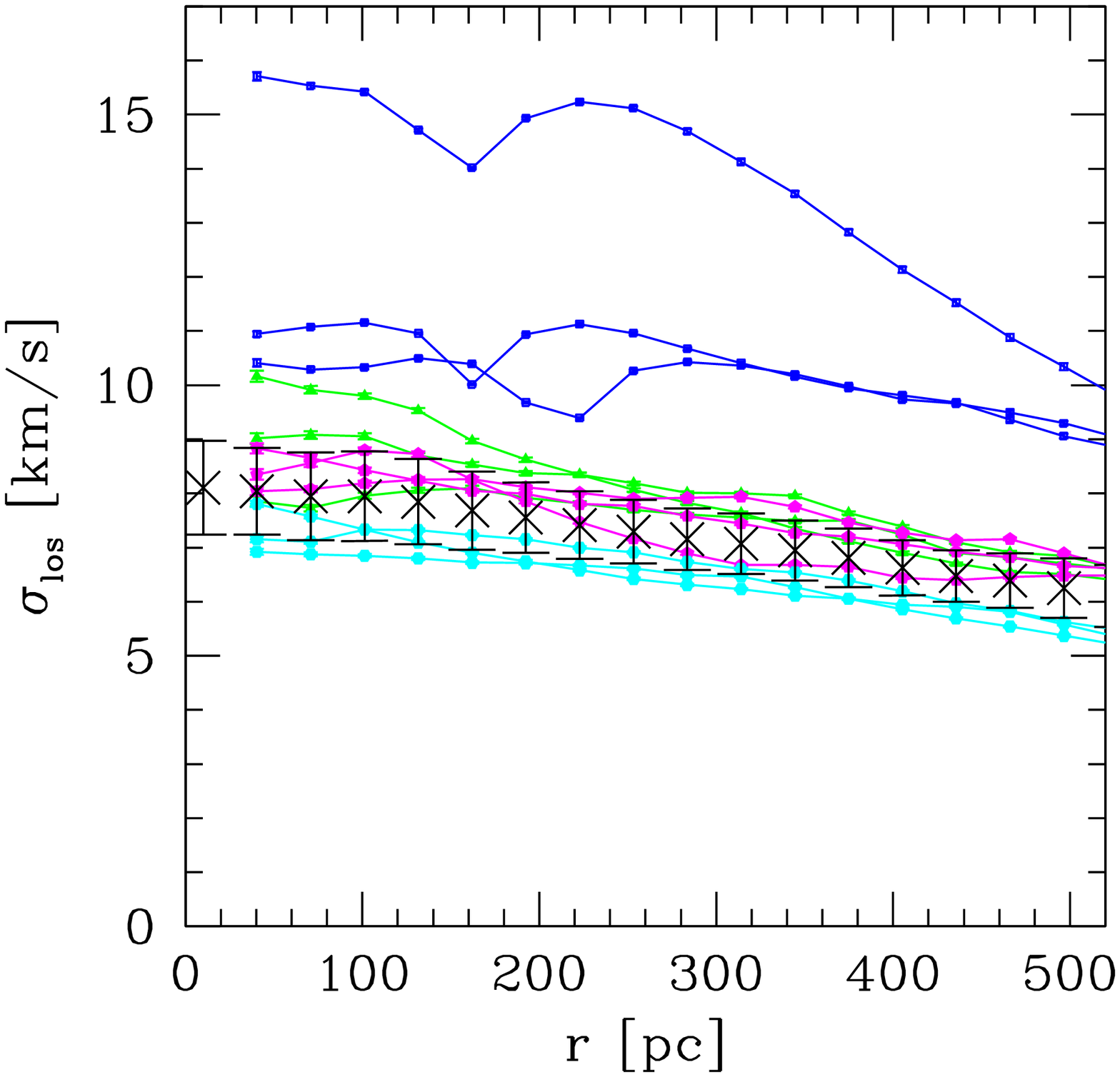}
    \epsfxsize=8.5cm
    \epsfysize=8.5cm
    \epsffile{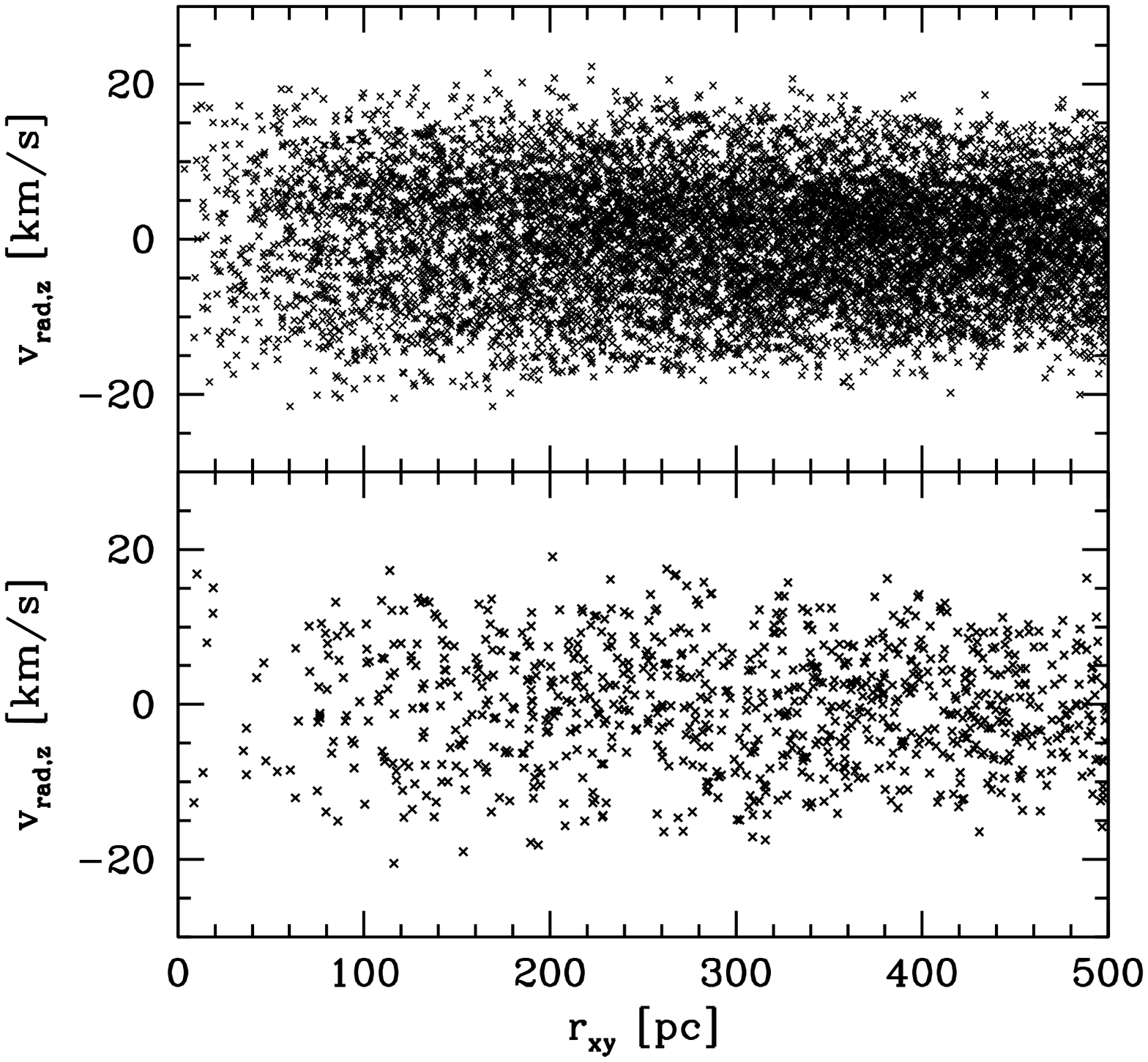}
    \caption{Left: 
      Line of sight velocity dispersion profiles along 
      the $x$-, $y$- and $z$-axis of our four fiducial models as open
      and closed symbols are shown (green = sim.1, blue = sim.2,
      magenta = sim.3, cyan = sim.4).  Error-bars for the single values
      are smaller than the symbol-size.  The mean values out of three
      simulations with 3 sight-lines each are shown as black crosses
      with error-bars (symbols as in Fig.~\ref{fig:sdens}).  
      Middle:
      Relative deviations of the mean velocity.  Shown as lines is the
      mean deviation of the three sight-lines, calculated in radial
      bins, of all four simulations (simulation 2, not used to derive
      the average, as dashed line) and the average thereof as the
      black crosses with error-bars.  The velocities are divided by the
      velocity dispersion calculated within 500 pc: $\sigma_{500}$.
      Right: 
      Radial velocity vs.\ radius, shown for model 1 and one
      sight-line. Top: one star in one hundred to show the richness of
      our simulation data; Bottom: one star in one thousand to mimic
      the best observational data available.}
   \label{fig:sigma}
 \end{center}
\end{figure*}

We calculate the 'clumpiness' of our model according to \citet{con03}:
\begin{eqnarray}
  \label{eq:clump}
  C & = & \frac{\sum_{\rm all pix} m_{\rm residual, pixel}}
  {\sum_{\rm all pix} m_{\rm original, pixel}} 
\end{eqnarray}
where the 'clumpiness' is defined as the sum of all positive fluxes
from the residuals divided by the sum of fluxes of the original data.
We use the IRAF routine ELLIPSE to calculate a smooth model,
which we then subtract from the original data, to obtain the
positive residuals. 

In our model the $C$ parameter is very low at $C = 0.045$.  This again
hints that the sub-structure we see in the brightness maps of our
model might be only visible due to our 'larger than reality' particle 
resolution.  The faint structures in our simulations are definitely real
and not due to noise.  But, once more, we have more particles in our
simulation than actual stars in the dwarf galaxy and definitely a
better particle resolution than observing only the brightest stars in
the dSph galaxy.  We therefore doubt that our formation scenario, which
produces very faint sub-structure in the surface-brightness map, will
be testable in the near future by purely photometric observations,
counting stars in pixels.  Instead we will show in a later section how
our model can be verified.

Looking at the smooth image we see that both the pitch angle and the
shape of the ellipses vary throughout the dwarf.  The pitch angle
varies between $-35.2$ to $-49.5$~degrees and the ellipticity shows
variations between $0.16$ and $0.47$, i.e.\ our model gets rounder in
the outer parts.  This again matches the ellipticities of the
classical dwarfs, which are on the order of $0.3$ \citep{irw95}.

We do not discuss the effects of the star clusters on the dark
matter halo because their total mass is very small compared with
the mass of the halo.  In our simulations the dark matter halo remains
cuspy during all the $10$~Gyr of evolution.  This is in accordance
with the recent work of \citep{cole11}, as our star clusters are
dissolving and therefore low-mass and low-density objects. 

\subsubsection{Model~2}
\label{sec:mod2}

Finally we should return to the model (number 2), which we have
singled out.  An inspection of the merger history of this model showed
that very early in the evolution of this model a few star clusters
merged in the very centre of the DM halo.  Here they were not subject
to dissolution due to the gas-expulsion.  In reality they might have
even retained their gas and formed a second generation of stars later
on.  In our model they lost their gas anyway (due to our
gas-loss prescription).  But instead of expanding to complete
dissolution they sat in the centre of the deep cuspy potential well of
the DM halo forming a dense nucleus.  If our formation process for
luminous components of galaxies is true not only for dSph but also for
larger, more massive objects, this by chance result of one
of our simulations could give a quite natural explanation why some
dwarf ellipticals have a nucleus and some not.  But this is
speculation beyond the scope of this paper.  In our case it only
depended on the random seed and therefore where the initial star
clusters were positioned.  As we see no nucleated dSph galaxies in the
Local Group, this could hint towards a much lower star formation
efficiency and a slower star formation rate in these dwarfs than
adopted in our fiducial model.

\subsection{Velocity Space}
\label{sec:svel}

\begin{table}
  \centering
  \caption{Velocity dispersion values and deviations from the mean
    velocity.  Column one gives the number of the realisation (again
    the asterisk marks the simulation we single out).
    $\sigma_{\rm los,0}$ is the velocity dispersion in the central pixel,
    $\sigma_{\rm los,0,fit}$ is the central value obtained by fitting a
    Plummer dispersion profile to the velocity dispersion data given in
    Fig.~\ref{fig:sigma}, $\sigma_{\rm los,500}$ is the velocity
    dispersion of all particles within a projected radius of $500$~pc
    and finally $\delta_{500}$ is the highest relative deviation of
    the mean velocity within $500$~pc, using a pixel size of $20$~pc.}  
  \label{tab:vel}
  \begin{tabular}{rrrrr} \hline
    Simulation & $\sigma_{\rm los,0}$ & $\sigma_{\rm los,0,fit}$ &
    $\sigma_{\rm los,500}$ & $\delta_{500}$ \\ 
    & [km\,s$^{-1}$] & [km\,s$^{-1}$] & [km\,s$^{-1}$] & \\ \hline
    Sim~1 & $7.5$ & $9.95 \pm 0.07$ & $9.0$ & $0.68$ \\
    $^{*}$Sim~2 & $11.3$ & $11.8 \pm 0.20$ & $14.5$ & $0.2$ \\
    Sim~3 & $8.4$ & $8.95 \pm 0.03$ & $7.0$ & $0.5$ \\
    Sim~4 & $7.3$ & $7.87 \pm 0.05$ & $8.3$ & $1.7$ \\ \hline
  \end{tabular}
\end{table}

\begin{figure*}
  \begin{center}
    \epsfxsize=5.8cm
    \epsfysize=5.8cm
    \epsffile{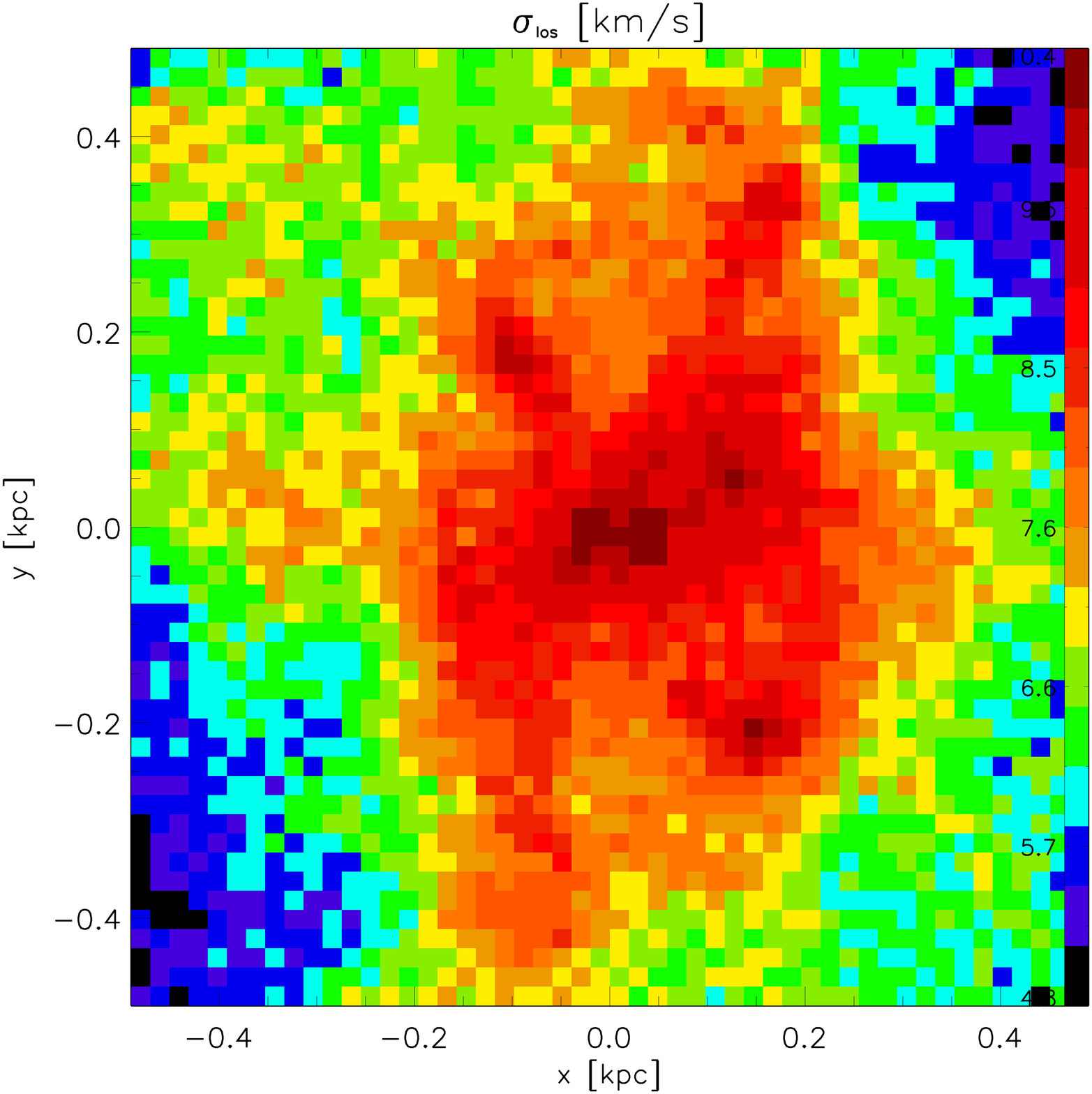}
    \epsfxsize=5.8cm
    \epsfysize=5.8cm
    \epsffile{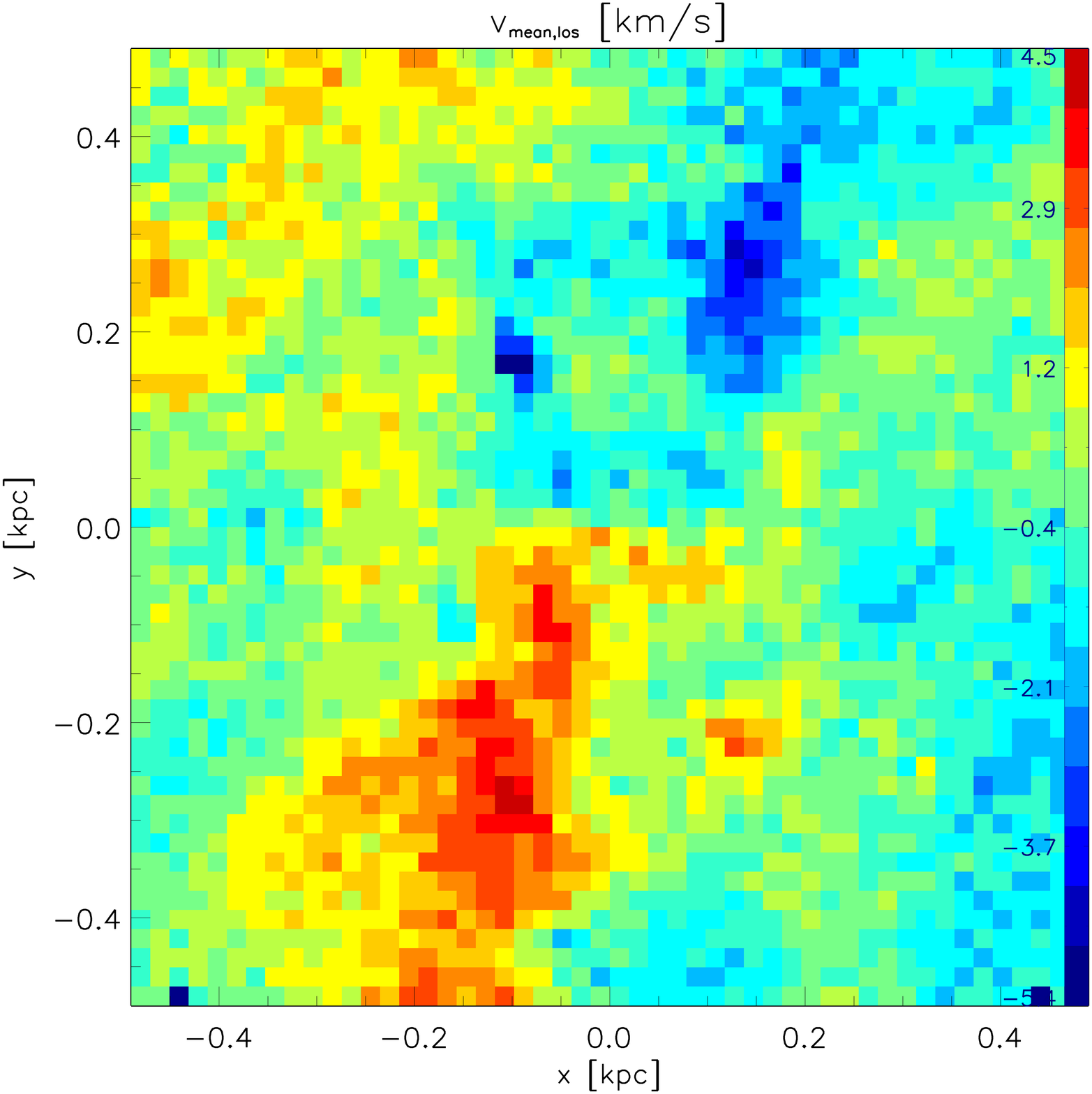}
    \epsfxsize=5.8cm
    \epsfysize=5.8cm
    \epsffile{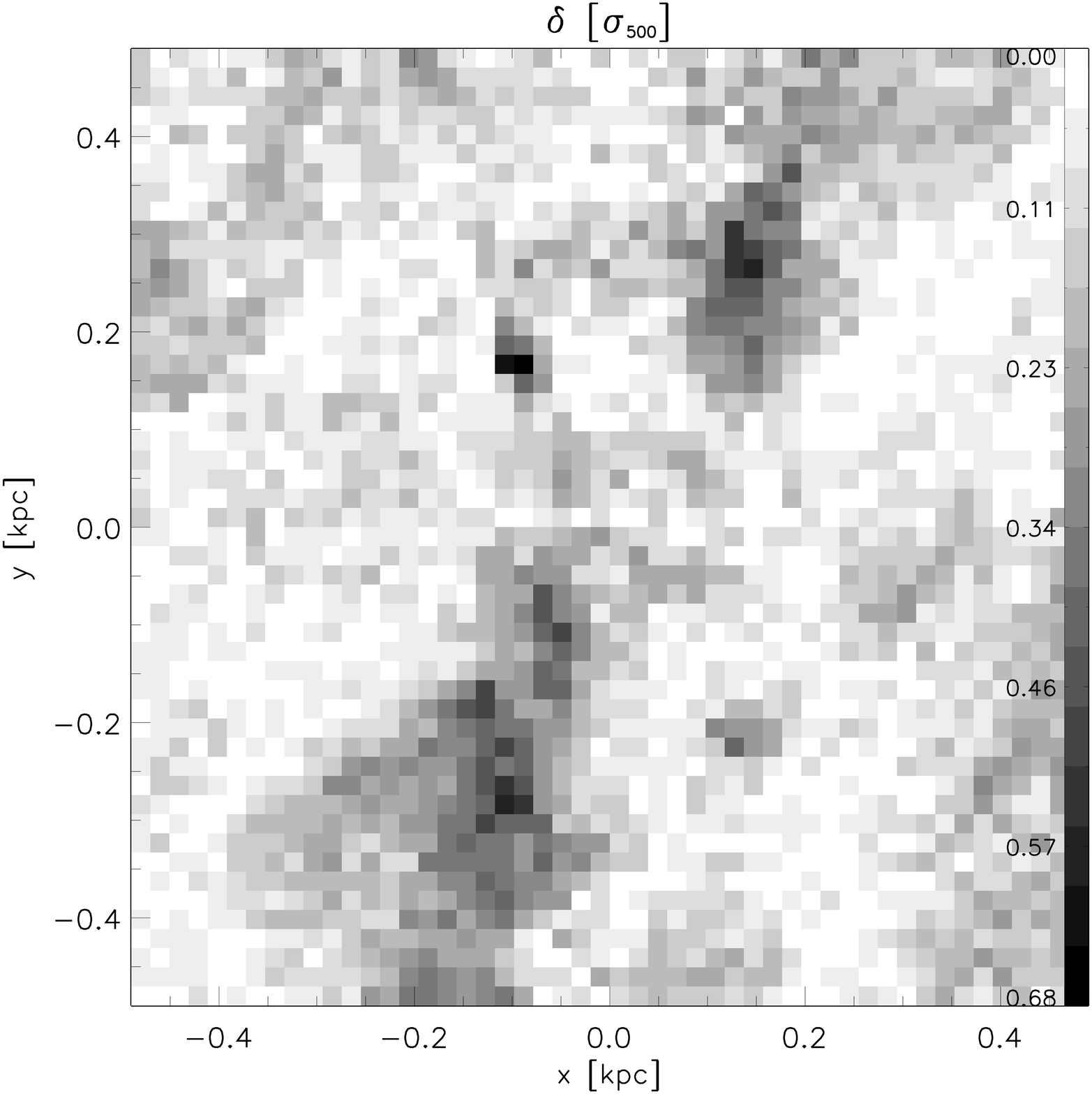}
    \caption{The 2D velocity structure of model 1. Shown is the
      central region out to $500$~pc with a resolution of $20$~pc per
      pixel. Left panel is a map of the velocity dispersion measured
      in each pixel.  Middle panel shows the mean velocity map
      calculated for each pixel separately and finally the right shows
      the scaled velocity deviation, which we call $\delta$-parameter
      (see main text for explanation).} 
   \label{fig:overview}
 \end{center}
\end{figure*}

In left panel of Fig. ~\ref{fig:sigma} we show the line-of-sight
velocity dispersion profiles ($\sigma_{\rm los}$) for our four
simulations and the crosses again show the average of three of our
realizations and the three Cartesian sight-lines.  Even though the
profiles seem to be quite below $10$~km\,s$^{-1}$ (we see a mean
central velocity dispersion of about $7.5$~km\,s$^{-1}$) the profiles
are almost flat, out to a radius of $500$~pc.  We see similar velocity
dispersion profiles in Leo~II and Sextans \citep{wal09}.  The lines
show the dispersion profiles of each sight-line and each realisation.
Here we see again mostly flat profiles but some profiles show wiggles,
dips or bumps.  While observers usually would have large error-bars
and would use a smooth curve to fit through the data, we argue that
those deviations from a smooth curve are real and due to our formation
process.  We also included error-bars for the data-points of the
single simulations, single sight-line data-points.  But these
error-bars are smaller than the actual size of our small symbols as
each data-point is based on 1,000 to 60,000 single velocities.
In fact these wiggles and bumps might be the only possible way,  
with the observations we have today, to verify our scenario.  Only 
realisation 2 shows a very high dispersion, which we discuss in a
separate section below.  Again we exclude this realisation from our
mean values.  The results of our four realisations are also given in
Tab.~\ref{tab:vel}.   

\begin{figure}
  \centering
  \epsfxsize=9cm
  \epsfysize=4.5cm
  \epsffile{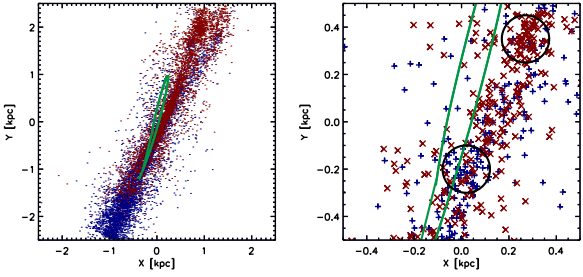}
  \caption{Orbit (green line) and particles of one dissolved star
    cluster.  Red crosses are particles with negative velocities, blue
    plus-signs have positive velocities.  At the right panel we see a
    blow up of the inner region and the circles show areas with mainly
    positive or mainly negative velocities.}
  \label{fig:s28}
\end{figure}

Finally, in the right panel of Fig.~\ref{fig:sigma} we show the radial
velocity of our particles of simulation~1 as function of their radius.
In the top panel we show one percent of our particles to emphasise how
well we are able to resolve the velocity space.  The lower panel shows
one particle in 1,000, which compares with the best available
observational data, present to date.  Again, in this low-resolution
plot, such as could be obtained from current observed data, we do not
see a strong sign of coherent motion, even though an inspection by eye
of the 'full' simulation data (upper panel) of Fig~\ref{fig:sigma}
would suggest some bi-modality in the very centre. 

If we use all stars within a radius of $500$~pc to compute an overall
velocity dispersion we get $\sigma_{\rm los,500,mean} =
9$~km\,s$^{-1}$ which resembles the values observed in classical dSphs 
as reported by \citet{mun05,wal07}.  This mean value of the velocity
dispersion velocity is calculated considering the lines-of-sight along
the three coordinates, because the sight-line of the dSph galaxy
towards us with respect to our formation scenario is unknown.    

%%%{\bf In case we use eq.~11 from \citet{wal09} to compute the total
%%%  dynamical mass within this $500$~pc radius, which corresponds to the
%%%  effective radius for this model, a velocity dispersion of
%%%  $\sigma_{500} = 9$~km\,s$^{-1}$ and the constant of $\mu =
%%%  580$~M$_{\odot}$\,pc$^{-1}$\,km$^{-2}$\,s$^{2}$ we obtain a mass of
%%%  $M(500{\rm pc}) = 2.35 \times 10^{7}$~M$_{\odot}$, more than double
%%%  the amount which is really present in our model
%%%  ($10^{7}$~M$_{\odot}$). 
%%%
%%%  If we repeat the exercise with the other realisations of the
%%%  fiducial model we get for model~3 a dynamical mass of $1.53 \times
%%%  10^{7}$~M$_{\odot}$ within an effective radius of $540$~pc.  In
%%%  reality the halo model used has only $1.12 \times
%%%  10^{7}$~M$_{\odot}$, i.e.\ we overestimate the halo mass by a factor
%%%  of $1.37$.  Model~4 has a dynamical mass of $1.80 \times 10^{7}$
%%%  within $450$~pc, $2.12$~times the DM-mass present.  Calculating a
%%%  mean value from the three realisations we get an overestimating
%%%  factor of $1.95 \pm 0.30$.
%%%
%%%  Only realisation~2, which we excluded shows completely different
%%%  values.  If we would use eq.~11 from \citet{wal09} on this model we
%%%  would result in an incredibly high overestimation of the DM-mass by
%%%  a factor of $15.0$.
%%%}   

In Fig.~\ref{fig:overview}, we show the dynamics of the
first realisation within $500$~pc as two-dimensional pixel maps with a
resolution of $20$~pc.  In the left panel we show the line-of-sight
velocity dispersion calculated for each pixel separately.  We see
regions of high velocity dispersion of more than $10$~km\,s$^{-1}$ in
the central area but also at the locations where the off-centre
density peaks are.  In high resolution the distribution of dispersions
is far from being smooth.  In fact we see even more sub-structure than
in the density plots.  While the DM cusp was able to erase most of the
structure in positional space, the structures in velocity space
survived.  We call these structures in velocity space which stem from
the formation process of the dwarf 'fossil remnants'.  Still we see
high values of velocity dispersion throughout the visible part
(surface brightness above $30$~mag\,arcsec$^{-2}$) of the dSph galaxy.  

In the middle panel we show the mean velocity of all particles within 
a pixel.  The figure shows clearly regions which seem to exhibit
coherent streaming motions with velocities differences of up to
$5$~km\,s$^{-1}$.  This means, even after $10$~Gyr of evolution we
still see the stars of dissolved star clusters, which were on rather
circular orbits in coherent motion.  The panel shows at least two
structures of opposing radial velocities.  A similar scenario was
proposed by \citet{kle03} to explain the velocity structure of UMi
(though in the plane of the sky and therefore with no detectable
offset in mean velocity).  

We define a $\delta$-parameter:
\begin{eqnarray}
  \label{eq:delta}
  \delta & \equiv & \frac{|{v_{\rm mean}}|} {\sigma_{\rm los,500}},
\end{eqnarray} 
where $v_{\rm mean}$ is the mean velocity measured in the radial bin
or later in a pixel and $\sigma_{500}$ is the velocity dispersion
measured within a projected $500$~pc radius.  This parameter allows us
to quantify how strong coherent motions in the object are.
The function $\delta$ corresponds to a re-normalization of the mean
velocities, which allows to see easily regions where the mean
velocities are distinct from zero.  Higher values of $\delta$
correspond to regions where streaming motions and/or rotation are
present.  

 While the value of the $\delta$-parameter depends on the strength
  of the velocity deviations in our model, the significance of those
  values, i.e.\ if they could be due to noise in the data-sample,
  depend on the number of velocities in the sample.  If we would apply
  a strict criterion to regard only the deviations which have
  $\delta$-values higher than possible $3 \sigma$-deviations (due to
  noise) as real, then we have to have values larger than $1.25$,
  $0.39$ and $0.14$ for sample sizes of $10$, $100$ and $1000$ stars
  respectively.  For more detailed values see table~1 in
  \citet{fel11}.

The right panel shows the $\delta$-parameter.  If no streaming 
motion is present we expect the $\delta$ values to be randomly
distributed without correlation.  As our high resolution plots have
more than 300 and up to 3000 particles in each pixel, the $\delta$
values we see of up to $0.68$ are definitely significant and not due 
to noise produced by low particle numbers.  

\begin{figure}
  \centering
  \epsfxsize=8cm
  \epsfysize=8cm
  \epsffile{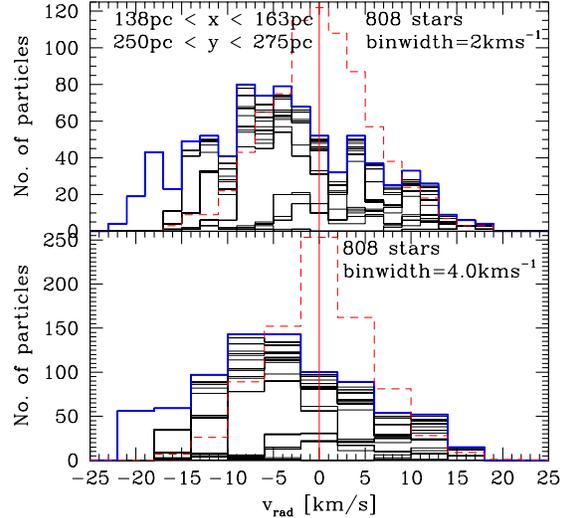}
  \caption{The right panel of Fig.~\ref{fig:overview} shows a pixel
    with a high $\delta=0.44$ parameter in the upper right part.  Here
    we show the radial velocity histogram of all 808 particles in this
    pixel as the solid (blue) histogram.  In the top panel we bin the
    data with $2$~km\,s$^{-1}$ bins and in the lower panel we show the
    same data with a bin-width of $4$~km\,s$^{-1}$.  For comparison, the  
    dashed (red) histogram shows a distribution of 808 randomly chosen 
    particles throughout the dwarf. The thin black histograms
      within the solid blue one show the cumulatively summed
      histograms of the particles from the $30$ different star
      clusters added one after the other}
  \label{fig:vdist}
\end{figure}

We speculate above that our fossil remnants could be related with star
clusters which dissolved on rather circular orbits.  In
Fig.~\ref{fig:s28} we show the particles of one of our dissolved
clusters.  Its initial orbit is shown as a green line on top of the
particles, which got spread out to much larger distances than the
original orbit.  Still they form a ring-like structure, where at the
upper end of the ring are more particles with negative velocities,
while at the lower end we see particles with positive velocities.  But
the 'ring' is too large for its ends to be the explanation of the two
places of high delta values we see in the right panel of
Fig.~\ref{fig:overview} (top right and lower left area).  If we zoom
in (right panel), we see that even though the ring is much larger
there are areas where the particles of this dissolved cluster have
predominantly positive or negative velocities (circles). These areas
coincide with the pixels of high delta values.

In Fig.~\ref{fig:vdist} we focus now on one pixel with a relatively
high $\delta$-value of $0.44$.  It is located in the upper right area
of Fig.~\ref{fig:overview}.  We plot the histogram of radial
velocities of all 808 particles in this pixel (taken from all
dissolved SCs, including the one we show in Fig.~\ref{fig:s28}).  The
top histogram, which has a good resolution of $2$~km\,s$^{-1}$, shows
that we have a complicated structure with several peaks and that the
majority of particles have negative velocities.  If we fit a Gaussian
to this data we obtain a mean value of $-4.7 \pm 0.3$~km\,s$^{-1}$ and
a $\sigma = 9.4 \pm 0.3$~km\,s$^{-1}$ (skewness: $0.202$;
kurtosis: $2.354$).  If we would have a 'normal' non-rotating
velocity distribution, the mean value has to be zero and the velocity
dispersion reflects the strength of the potential according to the
Jeans equation.  In reality, our histogram reflects many different
streams of particles from many different dissolved star clusters
rather than a smooth distribution of a uniform object in virial
equilibrium.  Therefore, the peak is shifted and the 'measured'
velocity dispersion is much higher than expected.  This is
  clearly visible if we look at the cumulative histograms of the
  different dissolved star clusters, which are shown as the thin,
  solid lines (black) in Fig.~\ref{fig:vdist}.  Larger areas without
  lines show where some star clusters deposit a lot of stars, i.e.\ we
  have a stream of stars from that particular cluster inside this
  pixel.  We also see that we do not have $30$ large areas inside this
  pixel, i.e.\ we do not find stars from every cluster in this pixel.

  This reflects the complicated velocity structure we have throughout
  the dwarf galaxy.  The particles of the different SCs are not
  distributed evenly and smoothly throughout the whole available
  phase-space.  Some of them still occupy distinct regions, reflecting
  the former orbit of the SC they came from.  

The better the resolution is, with which we view the object (smaller
areas, larger particle samples) the more visible are these structures.
In the lower panel we plot the same velocity sample with half the
resolution and the complicated multi-peak structure disappears, only
the shifted mean and the large $\sigma$-value remain.  

If we sample the same amount of velocities randomly from all over the
dwarf (red, dashed histograms in Fig.~\ref{fig:vdist})  the
distribution is indistinguishable from a normal Gaussian (skewness:
0.055; kurtosis: 3.305).  The mean value is centred on zero ($0.06 \pm
0.09$) and $\sigma = 5.35$~km\,s$^{-1}$ (Remember, we sample from the
whole object, i.e.\ also from the very outer parts showing low
dispersions; the velocity dispersion of all particles within $500$~pc
is still $\sigma_{500} = 9.0$~km\,s$^{-1}$). 

The result of these complicated velocity structures or 'fossil
remnants' as we dub them is that we could see regions with high
velocity dispersion and regions with deviations in mean velocity, if
we have high enough resolution, i.e.\ having both: small pixel size
and still enough stellar velocities per pixel.  If we do not have the
necessary resolution we still could see bumps and wiggles in the
velocity dispersion profiles and we might observe a velocity
dispersion that is enhanced relative to the equilibrium value for a
smooth model.  If we use the Jeans equations to estimate the mass of
the system we could then be biased to larger values of the mass.

\subsubsection{Model~2}
\label{sec:mod2-vel}

Again we have to discuss our singled out model separately.
Fig.~\ref{fig:sigma} shows that this model exhibits a 'cold' core with
a rising velocity dispersion outside the nucleus, reaching up to
$13.5$~km\,s$^{-1}$ and dropping again to values similar to the other
random realisations.  As said in the previous section, this simulation
exhibits a nucleus, which we do not find in the known dSph galaxies of
the MW.  But as expected for a nucleus inside a prominent DM halo we 
see a dip in the velocity dispersion.  The maximum in the
dispersion is due to the fact that in this particular model quite a
lot of the luminous mass got trapped in the nucleated centre instead
of being more distributed like the other models.   

\subsection{Comparing our model with dynamical observations of Fornax}
\label{sec:fornax}

\begin{figure}
  \centering
    \epsfxsize=4.1cm
    \epsfysize=4.1cm
    \epsffile{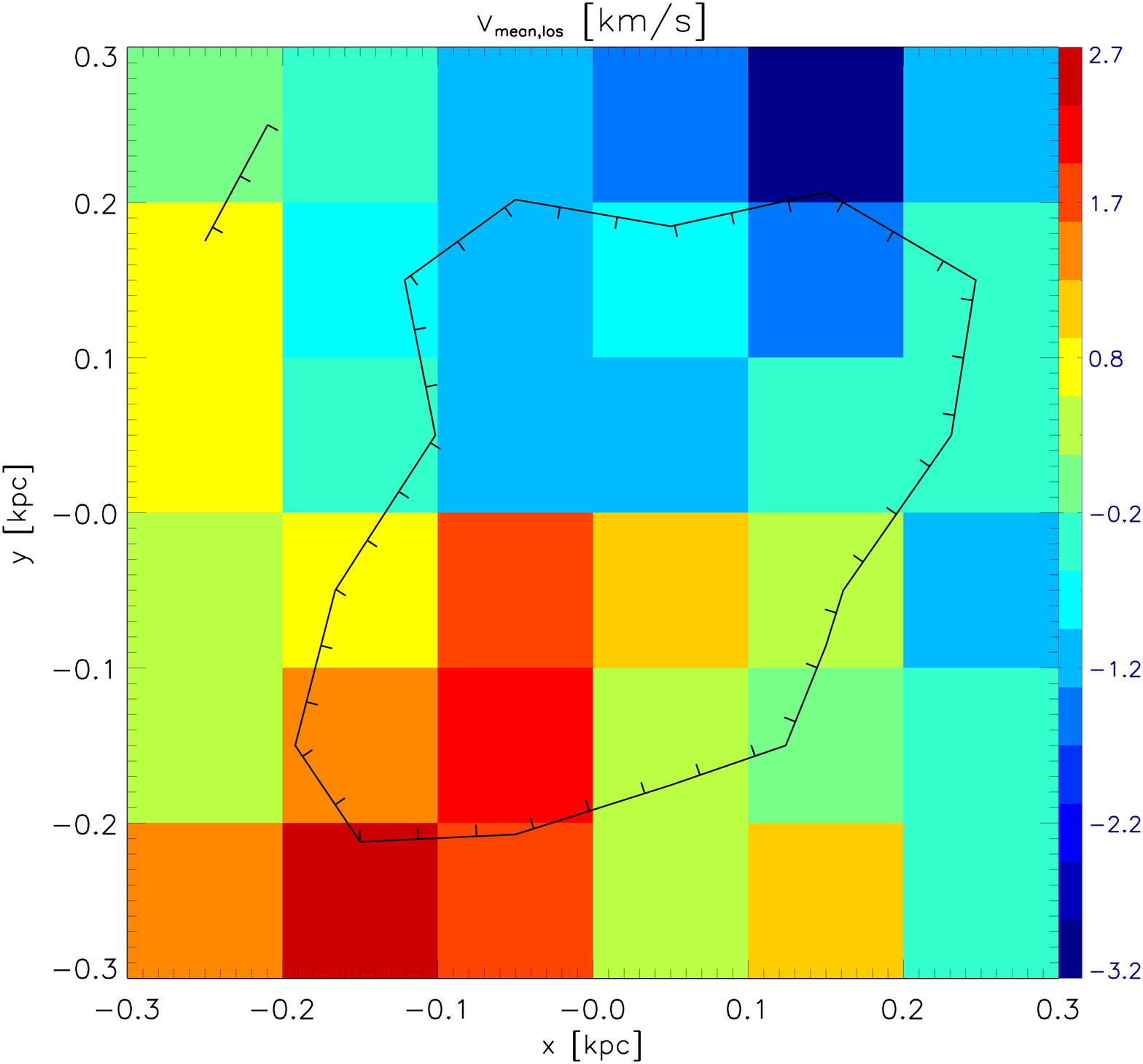}
    \epsfxsize=4.1cm
    \epsfysize=4.1cm
    \epsffile{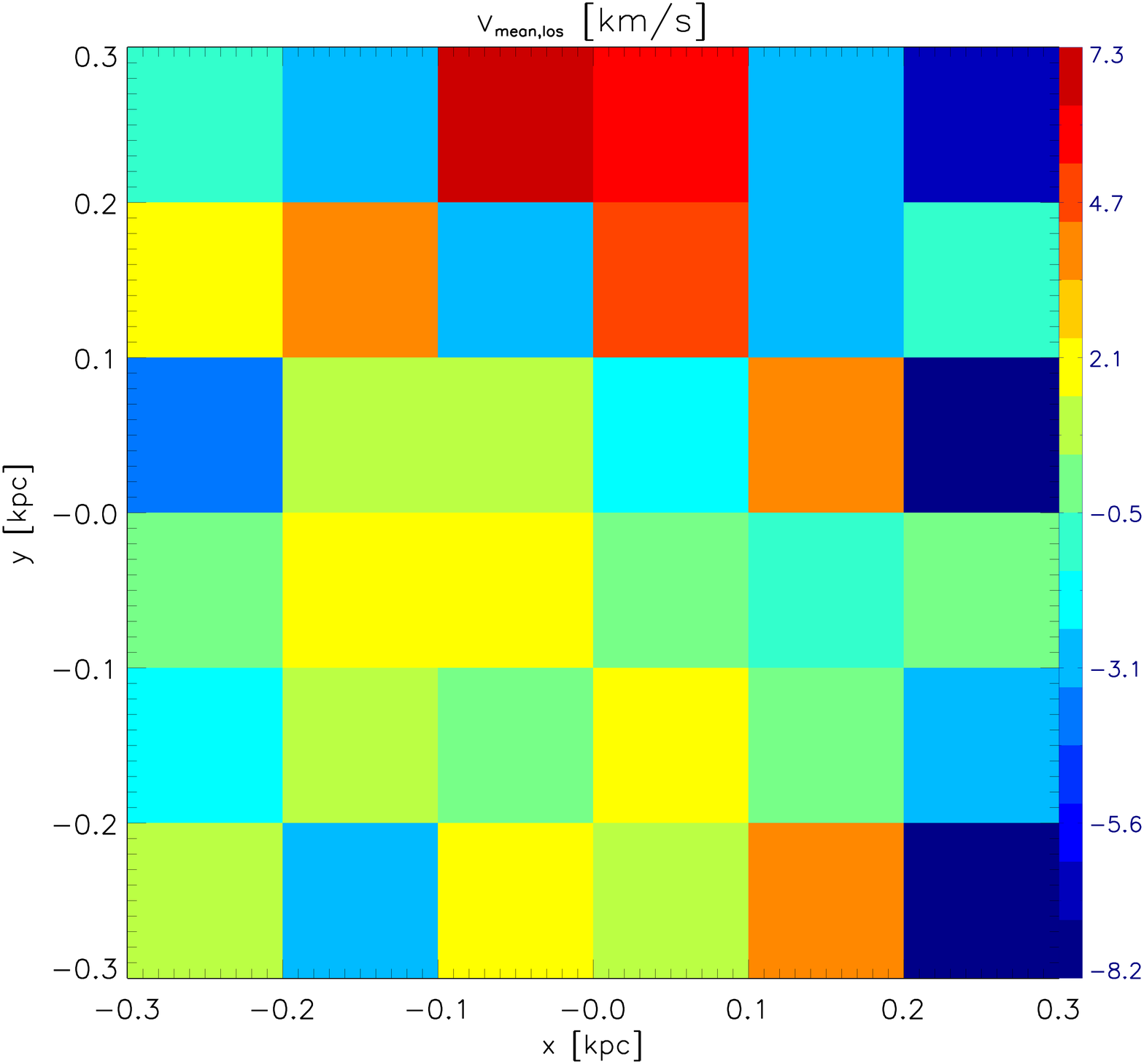}
    \epsfxsize=4.1cm
    \epsfysize=4.1cm
    \epsffile{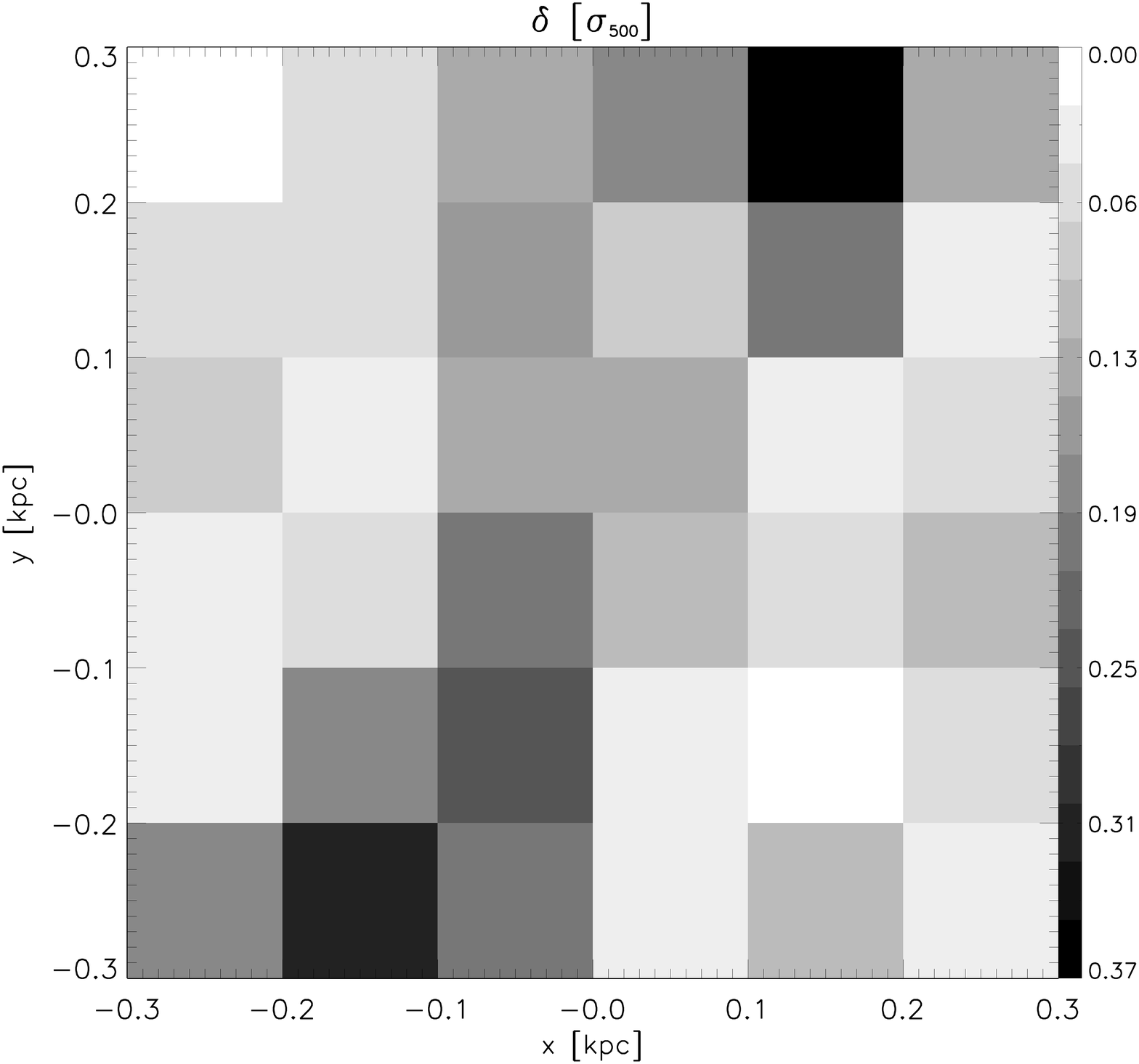}
    \epsfxsize=4.1cm
    \epsfysize=4.1cm
    \epsffile{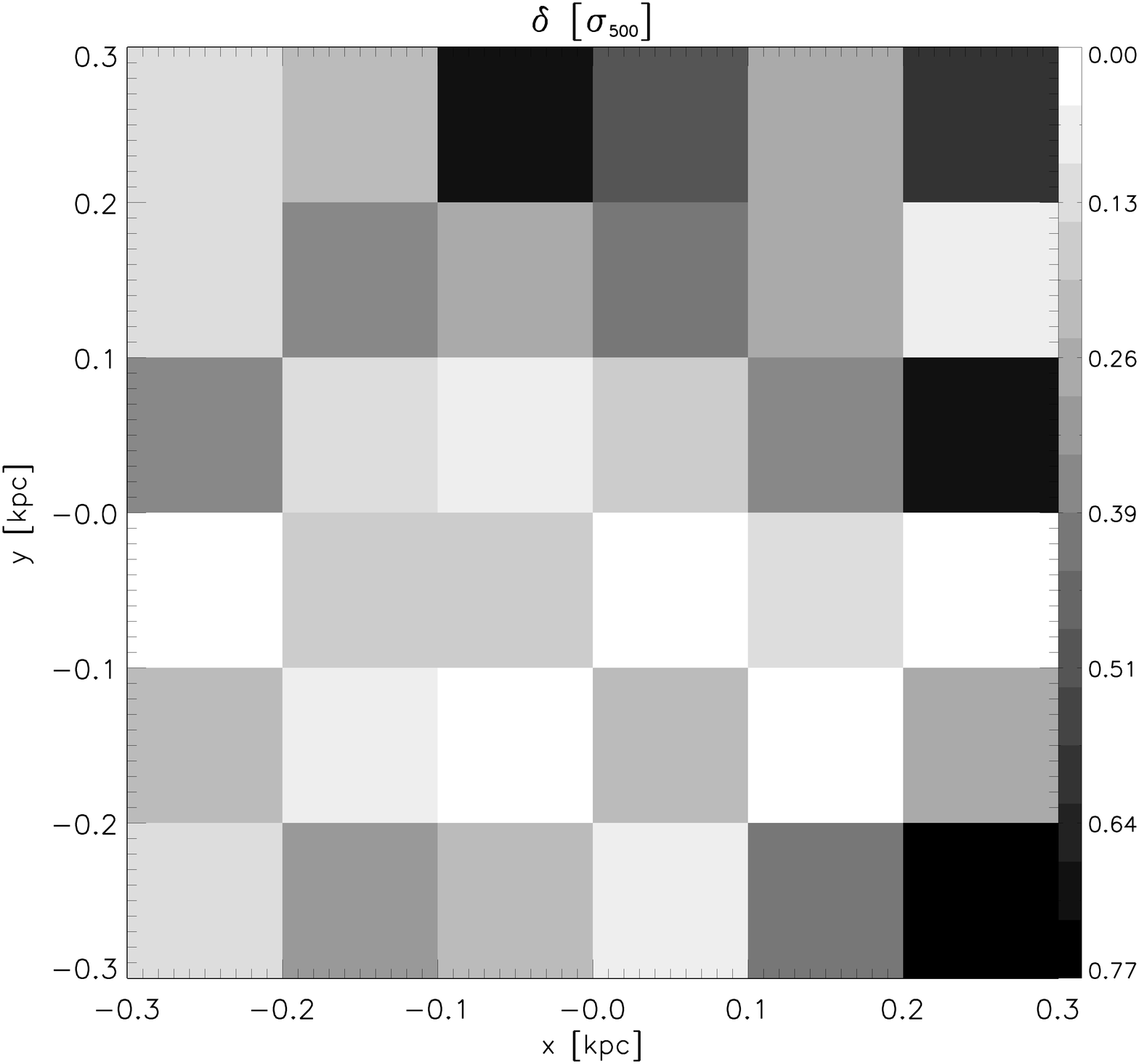}
  \caption{Contours with a very low resolution of $100$~pc per pixel
    showing mean velocity in the top row and $\delta$-parameter in the
    bottom row.  The left side is our model data and the right side is
    based on radial velocities measured in the Fornax galaxy.  The
    large velocity sub-structure in our model (top left panel),
    overlaid with black contours can easily be mistaken for an
    indication of rotation or a velocity gradient along the major
    axis.  See discussion in the main text.}
  \label{fig:lores}
\end{figure}

We try to compare our model with the available Fornax data in
Fig.~\ref{fig:lores}.  The top left panel shows the velocity
differences of the central part of our model with a resolution of
$100$~pc per pixel.  Overlaid is a contour of the surface brightness.
Within this contour the velocity structure could be mistaken as a
signal of rotation around the minor axis or a velocity gradient of
about $5$~km\,s$^{-1}$.  Underneath we see that the pixels showing the
strongest deviation have a $\delta$-parameter of about $0.5$.  Now
with our particle resolution we have more than 10,000 velocities in
each pixel and we can be sure that this velocity structure is real and
not due to noise.  Still, with this low resolution most signals of our
complicated velocity structure are erased.  The biggest structure is
just visible in the top panel due to a dark red and a dark blue pixel
and in the bottom panel through one dark pixel on either side of the
object.  We conclude therefore, that having only low resolution one
cannot see coherent motions (several pixels together showing the same
behaviour) but rather has to search for symmetric pixels with high
deviations. 

On the right hand side we try to produce a similar figure of the inner
part of Fornax.  Fornax has almost 2,000 radial velocities publicly
available.  In the region printed we have about 800 velocities.
Still, we have just around 20~velocities per pixel (minimum value: 8)
even with pixels this large.  There are pixels which show high
deviations in the mean velocity but first of all the strength of these
deviations is still within the possible noise (about a one-sigma
deviation from the system velocity, a $3 \sigma$-deviation due to
noise would result in delta values of $1.25$ for ten stars or $0.91$
for $20$~stars) and second we cannot detect any symmetry between the
pixels with high values.  So we have to conclude that the presently
available data of Fornax does not show any visible fossil remnant, as
we do not have sufficient data to disentangle possible sub-structure
in velocity space from the possible level of noise.  

Once again: even though our model and Fornax show deviations from the
systemic velocity of the same strength, in our model these deviations
are well above any noise and show symmetries, while Fornax lacks
sufficient data to enhance any structure, if present, above the
possible level of noise. If we would degrade our data to
  the same low resolution in pixel size {\it and} particle sampling,
  as available for Fornax, not a single fossil remnant would remain
  visible above the noise level.

\section{Discussion}
\label{sec:discus}

In this section we discuss in more detail the observational
detectability in real dSphs of the features we have seen in our
simulation. 

If we analyse Fig.~\ref{fig:smooth} in terms of detectability we
find that the visible sub-structure in the surface brightness
(bottom left panel) amounts to tiny differences in actual
mass-densities.  Within a $20 \times 20$~pc pixel the difference
between the dark red and red pixels amounts to a difference of about
$20$~M$_{\odot}$.  In our simulations we can easily resolve such mass
differences.  In reality, regarding the fact that most of this
mass-difference will be in faint low-mass stars, which are hard to
detect at the distance of the dSph galaxies even with modern
telescopes, these sub-structures will be below the detection limit.
Maybe the next generation telescopes like E-ELT, TMT or GMT will be
able to make those structures visible. 

But effects like slightly off-centre nuclei (Sextans), secondary
density peaks (Ursa Minor) or dents in the contours (Draco) (as seen
in \citet{irw95}) are now potentially explainable with our scenario as
low-resolution counterparts of a more complicated structure due to the
formation process.  

Looking at the velocity dispersion profiles of our fiducial models, we
find similar radial profiles to some of the classic dwarfs of the MW.
If we recall the profiles of the different realisations of our standard
model in Fig.~\ref{fig:sigma} we see that some of them show not only
an almost flat profile but also bumps or dents.  In our simulations
these deviations from a smooth curve are not due to noise (as we have
1,000 to 60,000 velocities in one bin) but are intrinsic deviations
stemming from our formation scenario.  They are hints to the fact that
a more complicated velocity structure exists in our model.  In reality
\citep{wal09} we see similar bumps in Carina, Leo~I and Sculptor, even
though the observational error-bars are still much larger than the
deviations. 

Even though some of the classical dwarfs have now more than
2,000 radial velocities, it is still not feasible to produce 2D
velocity maps of the dispersion or the mean velocity with a
sufficiently high resolution to detect 'fossil remnants', as we
see in Fig.~\ref{fig:lores} where we compare our simulations with
Fornax data set.  If with next generation telescopes several
  thousands of radial velocities more for each dSph galaxy would
  become available, significant sub-structure in velocity space might
  emerge above the expected level of noise.  In our fiducial model we
  found a maximum $\delta$-value of $0.4$, if we degrade our
  resolution to a pixel-size of $100$~pc.  In order for this feature
  to be detectable above the noise at a statistically significant
  level, we would need at least $100$ velocities per pixel (to be
  above a $3 \sigma$ noise level).  

Another interesting property of our model is that if you would
  draw an imaginary line along the maximum extension of our object in 
  surface brightness  (red/orange area in the left panels of
  Fig.~\ref{fig:smooth} or grey area in the middle panels of the same
  Figure, i.e.\ the projected major axis), the position angle of this
  line would be different to a similar line drawn through the maximum
  extension of high velocity dispersion (red/orange area in the left
  panel of Fig.~\ref{fig:overview}).  Furthermore, a line through the
  two regions of high mean velocity deviations (blue and red area in the
  middle panel of Fig.~\ref{fig:overview} or two main dark areas in
  the right panels of the same Figure), again would show a different
  position angle.  In other words, if one would fit ellipses of
  constant surface brightness and constant velocity dispersion to our
  data shown in Fig.~\ref{fig:smooth} and Fig.~\ref{fig:overview}
  (left panel), their position angles would differ.  Furthermore,
  searching for the position angle along which we see the highest
  velocity gradient in our data (middle panel of
  Fig.~\ref{fig:overview}), this angle would differ from the other
  two.  This is something which might only occur in our formation
scenario and therefore would be a measurable quantity.  Unluckily, a
perfect alignment would not exclude our scenario.  Just a perfect
alignment in all known dwarfs would render our scenario very
unlikely. 

One important ingredient of our models is the low SFE and therefore
the subsequent dissolution of the SCs.  Only with the dissolution of
the SCs we are able to lower the phase space density of the stars
sufficiently to obtain low luminosity objects like dSph galaxies.
Merging SCs without expansion always lead to compact objects, even
without the presence of a DM halo \citep{fel02}.  We see the
difference in our model~2, which had a chance merger early on in the
centre, thereby preventing the merged SCs from dissolving.  As such
objects as model~2 are not observed, we could conclude that the SCs
which formed the known dSph had even lower SFEs and slower SFR than
adopted in our model.

Last but not least, this paper only talks about our fiducial model,
i.e.\ one particular set of parameters that without much tweaking
results in an object which resembles (as shown in the section above) a
classical dwarf spheroidal galaxy in terms of appearance, scale-length
and velocity dispersion (without matching one particular dwarf of the
MW in every aspect).  A thorough investigation of the vast possible
parameter  space ($M_{500}$, $R_{\rm scale,halo}$, $R_{\rm scale,SC}$,
shape of the DM halo, distribution and virial ratio of the SCs,...)
will be dealt with in the follow-up papers we wish to publish.

Finally, one should bear in mind that after all, our model is still a
toy model making a lot of simplifications (all SCs form at the same
time, have the same mass and SFE, gas is lost entirely from the
system,...), some justified, some because we have no better knowledge.
Also here we have to investigate a much larger parameter space, which
we will do in the future.  A somewhat delayed formation of some of the
SCs might be a plausible explanation of the spread in metallicity we
see in dSph galaxies.  If not, a low SFR with low SFE also implies
that we will lack high-mass stars and strong super-nova explosions
\citep{wei10} which might be able to blow the remaining gas out of the
DM halo.  Retaining the gas could then lead to another epoch of star
formation later on.

\section{Conclusions}
\label{sec:conc}

In this paper we tested a possible scenario for the formation of dSph
galaxies with numerical simulations using the particle mesh code
{\sc Superbox} .  In our simulations we considered the evolution of $N
= 30$ star clusters within a dark matter halo with an NFW profile.  We
observe that these star clusters dissolve due to the low SFE $= 0.3$
assumed and form an extended luminous object which resembles a dSph
galaxy and after $10$~Gyr of evolution has a projected diameter of
about $1$~kpc.  

One major result to take away from our models is that even $10$~Gyr of
evolution is not enough to erase sub-structures stemming from our
formation process inside of the dSph.  These substructures are mainly
visible in velocity space.  Because of these sub-structures, i.e.\
coherent motions, the measurements of the velocity dispersion of the
object may be affected.  For this reason our fiducial model can mimic
the observations of a classical dSph galaxy even within a halo that
only has $M_{500} = 10^{7}$~M$_{\odot}$, a value somewhat lower than
the canonical values $M_{300}$ \citep{str08,wal09}  obtained by using
the measured velocity dispersion and assuming complete virial
equilibrium without coherent motions.     

The external structural remnants of the formation process in our model
are very faint, too faint to be observed in current data sets.  The
central structures which are within $500$~pc, have observable
luminosities comparable to the luminosity of the dSph galaxies of the
MW.  But the sub-structures stemming from the formation process
visible in the two-dimensional brightness distribution of our model
are only visible through our ultra-high particle resolution and are
currently unobservable, as long as we cannot see all the stars of the
dwarf.  If such structures are present, only future telescopes like
the E-ELT might be able to observe them. 

Still, those structures have low-resolution counterparts in off-centre 
nuclei, secondary density centres and oddly shaped contours, as
present in some of the dSph galaxies we know.

In velocity space the object has the overall properties of a typical
classical dSph galaxy.  We show that the velocity dispersion of this
object is around $10$~km\,s$^{-1}$ and that it has a flat velocity
dispersion profile in projection.  Again this resembles closely the
observations of the known classical dwarfs.

Besides we show the existence of streaming motions in this object even
after $10$~Gyr of evolution.  These streams are fossil remnants of the
dissolving star clusters considered in our simulations, mainly of
those with more circular orbits.  As mentioned above these kinematical
sub-structures only become visible in high-resolution pixel-maps.  The
current state of the art measurements in real dSph galaxies using up
to and more than 2,000 velocities in total is astonishing but do
still not give enough resolution to find those fossil remnants,
predicted by our model, in reality.  Our model showed one strong
fossil remnant, which might be detectable in real observations
possible today.  This structure is still visible if we
reduce our resolution to a resolution possible with present day
observational data.  It might be an explanation for the ring-like
structure suspected in UMi or it could be easily mistaken as a
velocity gradient throughout the luminous component, as for example
observed with UMa~II and Her.

It is important to note that these structures have 1D counterparts.
We see in the dispersion profiles dents and bumps.  In our simulations
these are not errors of the measurement but real features.  The same
behaviour is visible in some of the MW dSph galaxies.  In
observational papers these dents or bumps are smoothed over because
the data-points have much larger error-bars but according to our
scenario some of them could be real. 

Our simulations show that our new scenario for the formation of dSph
galaxies, joining two basic astronomical principles, indeed works.  We
are not only able to reproduce the observational data we have today,
but our models also provide observers with predictions for future
high-precision and high-resolution observations.
\\

{\bf Acknowledgments:}
PA wishes to thank W. Dehnen for his help with the NFW profiles. M. Metz for his support with Superbox. PA gratefully acknowledges support from a CONICYT PhD studentship,
MECESUP FMS0605, FONDAP no.~15010003, BASAL CATA grant PFB-06/2007,
FONDECYT no.~1100540 and FONDECYT no.~3130653.
MF acknowledges financial support through FONDECYT grant no.~1095092 
and BASAL.  MIW acknowledges the Royal Society for support.  RS is 
funded through a Comite Mixto grant and FONDECYT no.~3120135.

\label{lastpage}

\end{document}